\documentclass[a4paper,11pt]{article}
\usepackage{latexsym}	
\usepackage{amsmath}
\usepackage{amsthm}
 \usepackage{url}

\usepackage{xcolor}

\usepackage{ amssymb }
\def\ba{\begin{eqnarray}}
\def\ea{\end{eqnarray}}

\def\R{\hbox{\bf R}}

\def\ba{\begin{eqnarray}}
\def\ea{\end{eqnarray}}

\def\R{\hbox{\bf R}}
\def\lb{\label}
\def\be{\begin{equation}}
\def\ee{\end{equation}}

\def\C{\mathbb C}
\theoremstyle{plain}

\def\C{\mathbb C}
\def\R{\mathbb R}
\begin{document}

\title{Current induction and macroscopic forces for superconducting strings}
\author{Fernando Chamizo\thanks{Departamento de Matem\'aticas and ICMAT. Universidad Aut\'onoma de Madrid. 28049 Madrid, Spain,
fernando.chamizo@uam.es.} \and Osvaldo P. Santill\'an\thanks{Instituto de Matem\'atica Luis Santal\'o (IMAS), UBA CONICET, Buenos Aires, Argentina firenzecita@hotmail.com and osantil@dm.uba.ar.}}
\date {}
\maketitle

\begin{abstract}
Vortons are extended superconducting rings, which hypothetically may  play a role in cosmology and even may have significance in connection with cosmic rays of high energy.
Some of these objects are able to confine fermions which consequently become massless in the core of the object \cite{witten}, \cite{vorton3}. These fermions travel at light speed in the core and may generate a large current
without dissipation.  This raises interest about the generation mechanisms for these currents inside the defect. This question is analyzed here by studying the inverse photoelectric
effect for these objects namely, the absorption of a fermion with the consequent emission of a photon or a massive boson by the extended defect. Another motivation for the present work is that there exists a discussion in condensed matter
about the role of the bound spectrum in the macroscopic Magnus force that the vortex experiences in certain type of superfluids or superconductors. The discussion is about wether the main force comes from scattering of these fermions by the object or
by the effect of the environment on the bound states in the object, which may induce a spectral flow leading to an effective macroscopic force \cite{pelea1}, \cite{pelea2}, \cite{pelea3}. Without claiming that the results described here are conclusive in the context of condensed matter,  this work presents a comparison between these two effects for vortons interacting with a plasma of  fermions.
\end{abstract} 
\section{Introduction}
Among the topological defects that can be generated at early universe stages, string like objects are the most attractive at first sight. The reason is that monopoles or domain walls can overcome the critical density of the universe, the first due
to the super abundance of very light objects and the second due to the enormous energy contribution of the walls. The Standard Model does not include such objects. However, the possibility that its completion 
at large energies may accommodate them is not excluded. One interesting possibility are superconducting strings \cite{witten}, which in their fermionic version can confine fermions in their core
which travels at very large speeds without dissipation. This superconducting property, which is clear if the fermions are charged, has several consequences in cosmology. There exist some scenarios involving axions or $U(1)_{L-B}$ gauged models
which include this type of defects \cite{escenario1}-\cite{escenario8}. A particular set of these objects are vortons, which are superconducting rings. An initially small current is amplified \cite{vorton1}-\cite{vorton3} when the object makes dissipative contraction
until the centrifugal force of the current carriers compensates this contraction, thus rendering these objects stable. 

Although the existence of vortons is of vivid interest, there are some issues with the introduction of these objects in cosmology.  A conservative number of the scale of symmetry breaking allowed for these objects is $10^9$GeV, otherwise they may overcome the critical density of the universe at an unacceptable level \cite{vorton3}-\cite{vorton5}. This does not rule out these objects, but it excludes them as candidates for galaxy formation or to leave imprints in the cosmic microwave radiation. However, the assumption for those references is that these objects appear at the time of string formation, and this hypothesis was considered in \cite{ringeval}-\cite{ringeval2} by employing a Boltzmann formalism, allowing to these authors to enlarge the parameter space for these objects, see also \cite{vorton6}.
In addition, mechanisms that partially spoil the stability of these defects are of particular importance. Examples may be quantum tunneling, or modifications of the original Witten scenarios with further couplings leading to current leakage \cite{vorton9}-\cite{vorton99}.  An additional idea is to introduce electromagnetic corrections of the self-interactions of  the vorton with itself, which induce further decay channels \cite{gangui}. Still an interesting possibility is to consider fermionic superconducting strings, in which a zero mode fermion moving at a light speed inside the core. It is a possibility that, due to the way the corresponding mass generating scalar field couples, this mode ceases to be a zero mode when a phase transition occurs, thus the vorton stability is not ensured anymore \cite{davis}. 
Such phase transitions are likely to hold in the early universe,  and this may allows to enlarge  the $10^9$GeV value for the symmetry breaking scale. 

Despite these problems, there are further attractive properties of these objects, even when they do not violate the above mentioned symmetry breaking scale bound. It  is a possibility that vortons may generate cosmic rays of ultra high energy, above the GKZ energy \cite{rayos1}-\cite{rayos2}. In addition, if their collapse is gradual, it may have interesting applications in baryogenesis, as discussed for instance in \cite{rayos3} and references therein.

There is another apparently unrelated motivation for the present work, which comes from condensed matter physics and is related to vortices in superconductors and superfluids \cite{kopnin0}-\cite{experimento}. There exists in this context a discussion about the role of the bound states spectrum of the vortex on the forces the object experiences in these systems \cite{pelea1}, \cite{pelea2}, \cite{pelea3}. The reference \cite{pelea2} has proposed a force that does not depend on these bound states. On the other hand, based on the experimental data \cite{experimento} it was argued in \cite{pelea3} that the bound states living in the core of the vortex should have a macroscopic effect in the force these objects experience. The arguments of \cite{pelea3}
are based on  an effective picture in which  the equations for the bound state in a moving vortex resemble a Dirac equation with an electric field turned on. The application of anomaly arguments for this model \cite{kopnin23}, see also \cite{goldstone},  leads to the conclusion that there is a spectral flow of the energy levels, which may generate a macroscopic effect resulting into an effective vortex force. However, this point of view was argued in \cite{pelea1}, a reference which states that the scattering states are the most important for the force calculation. The reference \cite{kopnin4} suggests that, in the optical geometric limit, there is a precession phenomenom originated by the vortex rotation and Andreev reflection, and there is a tendency of the states to migrate from different angular momentum states, which is compensated by the vortex motion. Thus, the flow becomes oscillating and spectral flow does not really takes place. However, this reference also suggests that even though spectral flow is suppressed by these oscillations, it has some phenomenological consequence in the calculation of the coefficients of the Magnus and Iordanskii forces these objects experience. In any case, there is a debate about the role of the bound spectra for vortex physics in superconductors and superfluids.

As is suggested by the paragraphs given above, in the cosmological context the discussion is centered on vortons and their decays channels by assuming that the vorton is formed. However, for a vorton to mantain a superconducting current, it should absorb fermions somehow which become massless in the core. This capture effect is important since if, at the time of string formation, the corresponding cross section is suppressed, then only a little fraction of these defects become superconducting and this may be helpful in the context of the abundance problem. To the extent of the authors knowledge, there is little discussion in the literature about this capture and we would like to fill this gap. In addition, without claiming that the results to be presented here are conclusive in the context of condensed matter physics, a comparison between the macroscopic force on the vortex by the scattering of fermions and by this absorption process is presented. In other words,  the macroscopic effect of absorption of particles by the vortex, which is an effect due to the existence of bound state energies in the object, is analyzed for vorton scenarios. 

The present work is organized as follows. In \S2 the equations ruling fermions on the vortex are stated. Essentially they are the Dirac equation coupled to an electromagnetic field with cylindrical symmetry and to a massive field with angle dependence.
In \S3 the Fourier expansion of the bounded states is found for the case of general winding number in the scalar field via a synthetic complex analytical approach. This technique is also employed to fully compute the zero modes and to deduce a kind of Weyl's law for the energy levels.
In \S4 the cross section corresponding to the capture of a fermion is estimated, by working with certain analogy with the photoeffect. In \S5 it is discussed the Aharonov-Bohm scattering cross section to compare it to the result in the previous section. Finally, in \S6 it is studied the behavior of the corresponding cross section and the forces acting on the object as a function of the parameters of the problem.

\section{The Dirac equation for an abelian vortex}

The following description of the Dirac equation in presence of a vortex will be schematic, and the reader is referred to any QFT reference such as \cite{rubakov} for further details.

By gauge invariance arguments, as follows from formula (14.58) of  \cite{rubakov}, in presence of a gauge field $A_\mu$ and a complex scalar $\phi$, the Lagrangian density $\mathcal{L}=i\overline{\Psi}\gamma^\mu\partial_{\mu}\Psi-m\overline{\Psi}\Psi$ corresponding to the Dirac equation should be modified, under the Weyl representation, to
$$
\mathcal{L}= i\overline{\Psi}\gamma^\mu\,\text{diag}\big(D_\mu^-,D_\mu^+\big)\Psi
 -h\overline{\Psi}\,\text{diag}\big(\phi,\phi^*)\Psi.
$$
In the last expression the covariant derivative $D_\mu^\pm=\partial_\mu\mp\frac{1}{2}A_\mu$ was introduced and $h\in\R$ is the coupling constant. In the following the charge of the fermion will be taken as $e$, other options are possible
but they are not essential for our purposes.

To model a vortex along the $z$-axis, cylindrical symmetry is assumed  in the fields profiles except for an angular dependence in $\phi$ representing  a winding number~$n$. 
These fields are then parameterized as
$$
 A_\mu=
 \frac{B(r)}{r}
 (0,-\sin\theta,\cos\theta,0),
 \qquad
 \phi=ve^{in\theta}F(r)
$$
for some radial functions $B(r)$ and $F(r)$ and $v$ a real constant.
In principle $n$ is any integer (positive or negative). In this work special interest will be put  in the case $n=1$. The radius $r$ and the angle $\theta$
are the standard polar coordinates on the plane $xy$.

The symmetry of the problem and the static nature of the vortex suggest to look for solutions of the form $e^{i(kz-\omega t)}\Psi(x,y)$.
As shown in \S16.3 and \S14.3 of \cite{rubakov} the equations of motion characterize $\Psi$ as eigenstates
of the reduced operator  $\mathcal{D}\Psi=\lambda\Psi$ with\footnote{Note that in (16.45) there is a misprint with a wrong sign in $\phi$. However, the conclusions of that reference are, in our opinion, correct.} 
\begin{equation}\label{main_eq}
 \mathcal{D}=
 \begin{pmatrix}
  -i{\sigma}^1 D_1^- -i{\sigma}^2 D_2^- &-h\phi^*
  \\
  -h\phi & i\sigma^1 D_1^+ +i\sigma^2 D_2^+
 \end{pmatrix},
\end{equation}
where the reduced covariant derivative $D_j^{\pm}=\partial_j\pm \frac 12 i r^{-1}B(r)(-\sin\theta, \cos\theta)$ was introduced. 

By use of charge conjugation, it is enough to examine the case $\lambda\ge 0$ \cite{rubakov}.  
The operator $\mathcal{D}$ commutes with the anti-linear operator $\Psi=(\chi,\eta)\mapsto (\sigma^1\eta^*,\sigma^1\chi^*)$. This latter operator diagonalizes over $\mathbb{R}$
with eigenvalues $\pm1$. Then the eigenstates of \eqref{main_eq}  may be taken to satisfy $\eta=\pm \sigma^1\chi^*$. The quantities $\chi$ and $\eta$ of course, constitute a two component spinor.  The previous reduction to two degrees of freedom  allows to restrict ourselves to the first two coordinates of the eigenstate equation. After multiplying by $i\sigma^1$, they read
\begin{equation}\label{eq_1}
 (D_1^{-}+i\sigma^3D_2^{-})\chi -ihve^{-in\theta}F(r)\sigma^1\eta=i\lambda \sigma^1\chi
\end{equation}
where $\eta=\pm \sigma^1\chi^*$. The map $\Psi\mapsto i\Psi$ establishes a one-to-one correspondence between the spinors satisfying \eqref{eq_1} with $\eta=\sigma^1\chi^*$ and $\eta=-\sigma^1\chi^*$. Therefore the reduced Dirac equation
can be studied in the real space of the spinors satisfying the first condition and by then complexify it. In other words, the solutions of $\mathcal{D}\Psi=\lambda\Psi$ are of the form
\begin{equation}\label{spinor_c}
 \Psi=z_0\big(\chi, \sigma^1\chi^*)^T
 \qquad\text{with}\quad z_0\in\C.
\end{equation}
In connection with the possibility of taking $\eta =\pm \sigma^1\chi$, note that
changing the sign of $\eta$ is the same as changing the sign of $hv$. This implies that the positivity condition $hv>0$ can be freely assumed.

For simplicity, in the following discussion the profiles $B(r)$ and $F(r)$ will be taken as Heaviside step functions. This is an approximation for the true smooth functions appearing when solving the field equations for the corresponding Lagrangian (7.46) of \cite{rubakov} having boundary values $F(0)=B(0)=0$ and $F(\infty)=1$, $B(\infty)=n$ \cite{nohl}, which suggests
$$
 B(r) = n H(r-r_0)
 \qquad\text{and}\qquad
 F(r) = H(r-r_0)
$$
with $H(x)$ the Heaviside step function and $r_0$ a positive constant playing the role of the radius of the vortex. 

Note that, in the above approximation, the inner part of the vortex formally corresponds to disconnect the gauge and scalar fields putting $n=h=0$.

\section{The vortex eigenfunctions}

\subsection{A complex variable approach}
In the following, a complex variable approach for dealing with the reduced Dirac equations \eqref{eq_1} will be employed, which has the advantage of making
the presentation more compact. It will be shown that the resulting system of four first order equations can be reduced to two independent systems
of two first order equations. If the reader is not familiar or not confortable with the complex variable arguments given below, in the appendix there is a straightforward but brief
proof of the mentioned reduction.

In order to start, note that in \eqref{eq_1}, inside the derivatives in  $D_1^{-}+i\sigma^3D_2^{-}$, they appear the combinations $\partial_x\pm i\partial_y$.
The notation
$$
 \partial =\frac 12 (\partial_x-i\partial_y),
 \qquad z=x+iy=re^{i\theta},
$$
is standard in complex analysis, see for instance \S1.2 of \cite{duren}. The interest on the operator $\partial$ is that it annihilates anti-holomorphic functions and it is the standard derivative for holomorphic functions, so $$\partial z^\alpha = \alpha z^{\alpha-1}
\qquad \partial (z^*)^\alpha = 0.$$ This operator furthermore factorizes the Laplace operator in the form $4\partial\partial^*=4\partial^*\partial=\nabla^2$
and, as it will be shown later, acts as a raising order operator on Bessel like functions, which be fundamental for the present mathematical problem. 

With the notation given above and by taking into account that $\eta=\sigma^1\chi^*$ the equation \eqref{eq_1} reads
$$
 \begin{cases}
  2\Big(
  \partial^*-\frac{B(r)}{4z^*}
  \Big)
  \chi_1
  -i e^{-ni\theta}hvF(r)\chi_1^*=i\lambda \chi_2,
  \\[5pt]
  2\Big(
  \partial+\frac{B(r)}{4z}
  \Big)
  \chi_2
  -i e^{-ni\theta}hvF(r)\chi_2^*=i\lambda \chi_1.
 \end{cases}
$$
By the aforementioned property of $\partial$ with respect to holomorphic and anti-holomorphic functions,
for $\chi = (z^*/z)^{\alpha}\gamma$ the last equations become
$$
 \Big(
 \partial^*-\frac{\alpha}{z^*}
 \Big)\chi=
 \Big(\frac{z^*}z\Big)^{\alpha}
 \partial^*\gamma
 \qquad\text{and}\qquad
 \Big(
 \partial+\frac{\alpha}{z}
 \Big)\chi=
 \Big(\frac{z^*}z\Big)^{\alpha}
 \partial\gamma.
$$
If the last formula is applied in the case $\alpha= n/4$
the equations in the outer part of the vortex $r>r_0$ become
\begin{equation}\label{eq_out}
 2D\gamma -ihv\gamma^*=i\lambda\sigma^1\gamma
 \qquad\text{with}\quad
 D = \text{diag}\big( \partial^*, \partial\big).
\end{equation}
In the inner part of the vortex $r<r_0$, as noticed before, the equations correspond to $n=h=0$. This implies that they reduce to
\begin{equation}\label{eq_in}
 2D\chi=i\lambda\sigma^1\chi.
\end{equation}
It should be remarked that a branch cut appears in $\gamma$ if $n$ is odd since $(z^*/z)^{n/4}=e^{-in\theta/2}$.
Rigorously speaking, \eqref{eq_out} does not apply on that cut. This singularity of $\gamma$ will be compensated below to get a smooth spinor $\chi$.

Define the operator acting on a two component spinor $\beta$ by
$$
 \mathcal{C}\beta= -i\sigma^2\beta^*.
$$
It is a simple exercise to check that
$
  [i\sigma^1,1+\mathcal{C}]=0
$.
Trivially $[D,1+\mathcal{C}]=0$ and since the conjugate of $(1+\mathcal{C})\beta$ is $-i(1+\mathcal{C})\sigma^2\beta$, this means that the change $\gamma=(1+\mathcal{C})\beta$ in \eqref{eq_out} produces the equation $$2D\beta=i\lambda\sigma^1\beta+hv\sigma^2\beta.$$
This formula implies that in the outer part of the vorton
\begin{equation}\label{eq_outb}
 2\partial^*\beta_1=i(\lambda-hv)\beta_2,
 \quad
 2\partial\beta_2 =i(\lambda+hv)\beta_1
 \quad\text{with }\chi=e^{-in\theta/2}(1+\mathcal{C})\beta,
\end{equation}
and, correspondingly, in the inner part
\begin{equation}\label{eq_inb}
 2\partial^*\beta_1=i\lambda\beta_2,
 \qquad
 2\partial\beta_2 =i\lambda\beta_1
 \qquad\text{with}\quad\chi=(1+\mathcal{C})\beta.
\end{equation}
To assure the regularity of $\chi$ in the outer part,  the possible branch cut has to be cancelled by imposing that
\begin{equation}\label{ddelta}
 e^{i\theta\delta}\beta
 \quad \text{is regular with}\quad
 \delta =
 \begin{cases}
  0&\text{if $n$ is even},
  \\
  1/2&\text{if $n$ is odd}.
 \end{cases}
\end{equation}
Since $4\partial\partial^*=\nabla^2$,  then the task of solving \eqref{eq_outb} reduces to finding $\beta_1$ such that
\be\lb{boils}
 \nabla^2\beta_1=-(\lambda^2-h^2v^2)\beta_1,
 \ee
 and, after that, by taking
$$
 \beta_2= -2i(\lambda-hv)^{-1}\partial^*\beta_1.
$$
The same argument applies to \eqref{eq_inb} by putting $h=0$.

\subsection{Bound states and energies}

According to the previous paragraph, in particular equation (\ref{boils}), the problem boils down to solve an equation of the  Helmholtz type $$\nabla^2 u=-\mu^2u.$$ It is well known, see for instance V.5. of \cite{CoHi}, that
$e^{i\nu\theta}Z_\nu(\mu r)$ is a solution when $Z_\nu$ satisfies the Bessel equation. In the classical theory $\nu\in \mathbb{Z}$, but here  the half integral case has to be considered as well due to the branch cut for $n$ odd. The standard basis of the Bessel functions is
given by $\big\{J_\nu(x), Y_\nu(x)\big\}$
with $Y_\nu(x)$ having a singularity at the origin and both functions decaying as $x^{-1/2}$. If $\mu^2<0$, the parameter $\mu$ is pure imaginary and, although one may consider
$\big\{J_\nu(ix), Y_\nu(ix)\big\}$ as a basis, the standard basis is
$\big\{K_\nu(x), I_\nu(x)\big\}$
with $$K_\nu(x)\sim e^{-x}/\sqrt{x}\qquad I_\nu(x)\sim e^{x}/\sqrt{x},$$
when $x$ is large.
By employing that $2\partial^*=e^{i\theta}\big(\partial_r+ir^{-1}\partial\theta)$ or by expressing $r$ and $\theta$ in terms of $z$ and $z^*$, this together with the recursion formula (8.472.2) of reference \cite{GrRy} leads to
\begin{equation}\label{recur}
 2\partial^*\big(e^{i\nu\theta}Z_\nu(\mu r)\big)
 =-\mu e^{i(\nu+1)\theta}
 Z_{\nu+1}(\mu r).
\end{equation}
After this brief summary about Bessel functions, the task is to solve \eqref{eq_inb} and \eqref{eq_outb}. The unknown $\beta_1$ can be expanded in the inner part into Fourier series $\beta_1=\sum_k e^{ik\theta}f_k(r)$. As the case under study is $\lambda^2>0$, then the previous considerations show that $f_k$ is a multiple of $J_k(\lambda r)$ and from this, it is found that
$$
 \beta =
 \sum_{k\in\mathbb{Z}}
 \begin{pmatrix}
  a_k
    \\
  ia_{k-1}
 \end{pmatrix}
 e^{ik\theta}J_k(\lambda r)
 \quad\text{and}\quad
 \chi =
 \sum_{k\in\mathbb{Z}}
 \begin{pmatrix}
  A_k
    \\
  iA_{k-1}
 \end{pmatrix}
 e^{ik\theta}J_k(\lambda r)
 \qquad
 \text{for }r<r_0.
$$
The operator $1+\mathcal{C}$ does not change the form of the expansion because $\chi$ and $\beta$ satisfy the same equations namely, \eqref{eq_in} and \eqref{eq_inb}.

Instead, in the outer part, as $e^{i\theta\delta}\beta$ has to be regular, the Fourier expansion to consider is
$\beta_1=\sum_k e^{i(k-\delta)\theta}f_k(r)$.
For $\lambda^2-h^2v^2>0$ there are not bound states since $J_{k-\delta}$ and $Y_{k-\delta}$ do not decay quick enough. If instead $\lambda^2-h^2v^2<0$ one is driven to take $f_k$  as a $K_{k-\delta}$ Bessel function. Hence
\begin{equation}\label{Kexp}
 \beta =
 \sum_{k\in\mathbb{Z}}
 \begin{pmatrix}
  s_{-}b_k
    \\
  -is_{+}b_{k-1}
 \end{pmatrix}
 e^{i(k-\delta)\theta}K_{k-\delta}(\lambda_{hv}r)
 \qquad
 \text{for }r>r_0
\end{equation}
where
$s_{\pm}=\sqrt{hv\pm\lambda}$ and
$\lambda_{hv}=s_{+}s_{-}= \sqrt{h^2v^2-\lambda^2}$.

The condition imposing that the solutions of \eqref{eq_outb} and \eqref{eq_inb}
agree on the boundary $r=r_0$ is
$$
 e^{-in\theta/2}(1+\mathcal{C})
 \sum_{k\in\mathbb{Z}}
 \begin{pmatrix}
  s_{-}b_k
    \\
  -is_{+}b_{k-1}
 \end{pmatrix}
 e^{i(k-\delta)\theta}K_{k-\delta}(\lambda_{hv}r_0)
 =
 \sum_{k\in\mathbb{Z}}
 \begin{pmatrix}
  A_k
    \\
  iA_{k-1}
 \end{pmatrix}
 e^{ik\theta}J_k(\lambda r_0).
$$
Comparing the Fourier coefficients of both sides, linear conditions are obtained.

Let us work out the details for $n=1$ (hence $\delta=1/2$) to get the energy equation. In this case, after applying $1+\mathcal{C}$,  it is deduced equating the $k-1$ Fourier
coefficients
\begin{equation}\label{compc}
 \begin{pmatrix}
  s_{-}b_k-is_{+}b_{-k}^*
    \\
  -is_{+}b_{k-1}+s_{-}b_{1-k}^*
 \end{pmatrix}
 K_{k-1/2}(\lambda_{hv}r_0)
 =
 \begin{pmatrix}
  A_{k-1}
    \\
  iA_{k-2}
 \end{pmatrix}
 J_{k-1}(\lambda r_0).
\end{equation}
The first coordinate expresses $A_{k-1}$ in terms of $X=b_k$, $Y=b_{-k}^*$ and the second coordinate after the shift $k\mapsto k+1$ gives an independent expression for $iA_{k-1}$. The condition to be both compatible is
\begin{equation}\label{hom1}
 \frac{K_{k-1/2}(\lambda_{hv}r_0)}{J_{k-1}(\lambda r_0)}
 \big(s_{-}X-is_{+}Y\big)
 =
 -i
 \frac{K_{k+1/2}(\lambda_{hv}r_0)}{J_{k}(\lambda r_0)}
 \big(-is_{+}X+s_{-}Y\big).
\end{equation}
On the other hand, conjugating the first coordinate and changing $k\mapsto -k$ the value of
$A_{-k-1}^*$ is expressed as a linear combination of $X$ and $Y$, by use of  the symmetries $J_{-n}=(-1)^nJ_n$ and $K_{\nu}=K_{-\nu}$ which follow from the formulas (8.404.2) and (8.486.16) of \cite{GrRy}. Another formula for $A_{-k-1}^*$ follows conjugating the second coordinate and changing $k\mapsto 1-k$. The compatibility of both formulas imposes this time
\begin{equation}\label{hom2}
 \frac{K_{k+1/2}(\lambda_{hv}r_0)}{J_{k+1}(\lambda r_0)}
 \big(s_{-}Y+is_{+}X\big)
 =
 -i
 \frac{K_{k-1/2}(\lambda_{hv}r_0)}{J_{k}(\lambda r_0)}
 \big(is_{+}Y+s_{-}X\big).
\end{equation}
No more equations can be deduced from \eqref{compc} involving $b_k$, $b_{-k}$, $A_{k-1}$ or $A_{-k-1}$. Therefore, it is concluded that the existence of nontrivial solutions of the homogeneous linear system \eqref{hom1}, \eqref{hom2} gives rise to a bound state. It happens exactly when the determinant vanishes, leading to the energy equation
\begin{equation}\label{eneq}
 \lambda_{hv}
 \Big(
 \frac{K_{k+1/2}}{K_{k-1/2}}
 J_{k-1}
 -
 \frac{K_{k-1/2}}{K_{k+1/2}}
 J_{k+1}
 \Big)
 J_k
 =
 \lambda
 \big(
 J_k^2+J_{k-1}J_{k+1}
 \big)
\end{equation}
where for the sake of simplicity the arguments $\lambda_{hv}r_0$ and $\lambda r_0$ have been omitted in the $K$ and $J$ Bessel functions, respectively.

Just for illustration, consider the case $k=1$, $hv=5$ of the previous analysis with $r_0=0.34845416$. This value has been chosen to force the energy equation to have the solution $\lambda=4$. Introducing these data in \eqref{hom1} and \eqref{hom2} the resulting underdetermined linear system shows that $(b_1,b_{-1})$ is any multiple of $(1, 7.6709342\,i)$. Taking this as its value and recalling that both sides of \eqref{hom1} equal to $A_0$ and both sides of \eqref{hom2} to $A_{-2}^*$. It is deduced
$A_0=-16.637090$
and
$A_{-2} = 19.134381\,i$.

Substituting these numerical values in the formulas for the Fourier expansions and setting the rest of the $b_j$ and $A_j$ as zero, it is obtained that the $\chi$-part of the bound state corresponding to $\lambda=4$ is given by
\[
 \chi=
 \begin{pmatrix}
  19.134381i \, J_{2}(4 r) e^{-2i\theta} - 16.637090 \, J_{0}(4 r)
  \\
  -16.637090i \, J_{1}(4  r) e^{i\theta} + 19.134381 \, J_{1}(4  r) e^{-i\theta}
 \end{pmatrix}
 \qquad\text{for }r\le r_0.
\]
and, substituting $K_{1/2}$ and $K_{3/2}$ by their formulas as elementary functions,
\[
 \chi=
 r^{-3/2}e^{-3r}
 \begin{pmatrix}
  (3.3798938 \, r
  + 1.1266313) \, ie^{-2i\theta}
  - 15.928492 \, r
  \\
  (-7.7215014 \, r
  - 2.5738338) ie^{i\theta}
  + 17.375694 \, r e^{-i\theta}
 \end{pmatrix}
 \qquad\text{for }r\ge r_0.
\]
The constants have been rounded to $8$ significant digits. Recall that the spinor is completed with the $\eta$-part $\eta=\sigma^1\chi^*$. Note the exponential decay in the last formula and the continuous definition.

\subsection{The zero mode space}
A particular physical important set of eigenvalues and eigenvectors are the zero modes, that is, solutions corresponding to $\lambda=0$. This is due to the fact
that these modes are the ones that may generate large currents inside the extended object.
According to \eqref{eq_in} and \eqref{eq_out}, the zero modes correspond to solutions of
\begin{equation}\label{zmodes}
 \begin{cases}
   2
 \partial^*
 \chi_1=0,
 \quad\qquad
   2
 \partial
 \chi_2=0
 &\text{if }r<r_0,
 \\
 2
 \partial^*
 \gamma_1
 =i hv\gamma_1^*,
 \quad\;
  2
  \partial
  \gamma_2
  =i hv\gamma_2^*
 &\text{if }r>r_0,
 \end{cases}
\end{equation}
where $\chi = e^{-in\theta/2}\gamma$ for $r>r_0$ and  $\chi$ is continuous through the boundary.
The equations are decoupled and are transformed into each other under the change $\gamma_2$ into $-i\gamma_1^*$. Then the attention is focused on solving
$$
 \partial^*(e^{-in\theta/2}\gamma_1) = 0\quad\text{if }r<r_0,
 \qquad
 2\partial^* \gamma_1= ihv\gamma_1^*\quad\text{if }r>r_0.
$$
The first equation is satisfied for any $e^{-in\theta/2}\gamma_1$ holomorphic.
On the other hand, writing $\gamma_1(\mathbf{r})=(1-i) f(hv\mathbf{r})$ the second equation is $2\partial^* f= -f^*$ for~$r>hvr_0$. It follows the characterization of the real space of the $\chi$-part of the zeros modes\footnote{Recall that $\eta=\sigma^1\chi$.}
as the linear space generated by
\begin{equation}\label{zmspace}
 \left\{
 (1-i)e^{-in\theta/2}
 \begin{pmatrix}
  f_-(hv\mathbf{r})
  \\
  f_{+}^*(hv\mathbf{r})
 \end{pmatrix}
 \,:\,
 f_{\pm}\in S^\pm
 \right\}
\end{equation}
where
$S^\pm$ is the (real) space of functions $f$ with $e^{\pm in\theta/2}f$ normalizable and continuous on $r=hvr_0$ such that
$$
 2\partial^*f=-f^*\quad\text{in }r>r_1=hvr_0
 \quad\text{ and }\quad
 e^{\pm in\theta/2}f\quad\text{holomorphic in }r<r_1.
 $$
By conjugating the equation and by substituting it into itself, it is found that $\nabla^2f=f$  which corresponds to the case $\mu=i$ of the Helmholtz equation leading as before  to
$$f=\sum_{k\in\mathbb{Z}} c_ke^{i\theta (k-\delta)}K_{k-\delta}(r).$$ The recursion formula \eqref{recur} implies that $2\partial^*$ acts on $e^{i\theta (k-\delta)}K_{k-\delta}(r)$ just shifting $k$ into $k+1$, hence
$2\partial^*f=-f^*$
imposes $c_k^*=c_{2\delta-1-k}$ and is equivalent to it. This is clearly seen by taking into account that $K_{\nu}=K_{-\nu}$ , as follows from formula (8.486.16) of reference \cite{GrRy}.

Form the above discussion, it is concluded that
$f$ in $r>r_1$, the outer part, belongs to the real space generated by
\begin{equation}\label{zmout}
 \big\{H_\nu^{\text{o}}+H_{-1-\nu}^{\text{o}},iH_\nu^{\text{o}}-iH_{-1-\nu}^{\text{o}}\big\}_{\nu\in \delta+\mathbb{Z}}
 \quad\text{with}\quad
 H_\nu^{\text{o}}=K_{\nu}(r)e^{i\nu \theta}
\end{equation}
where $\delta$ is as in \eqref{ddelta}.
By the symmetry $\nu\mapsto -1-\nu$ it can be freely assumed that $\nu\ge -1/2$ and in this range the previous set is linearly independent except for $iH_\nu^{\text{o}}-iH_{-1-\nu}^{\text{o}}=0$ when $\nu=-1/2$, which only appears for $n$ odd ($k=0$, $\delta=1/2$).

Now, an analytic function is uniquely determined by its values on a curve. This means that the only holomorphic function in $r\le r_1$ matching $A_1 e^{im_1\theta}+A_2e^{im_2\theta}$ on $r=r_1$, with $A_1,A_2\ne 0$, is
$$A_1(z/r_1)^{m_1}+A_2(z/r_1)^{m_2},$$ 
which forces $m_1$ and $m_2$ to be nonnegative integers. So, the necessary and sufficient condition for the elements in \eqref{zmout}  to belong to $S^\pm$ is $\nu\pm n/2, -(\nu+1)\pm n/2\in\mathbb{Z}_{\ge 0}$.
As   $\nu\ge -1/2$ both numbers are integers and differ in a nonnegative integer. Therefore it is enough to require $\pm n/2\ge \nu+1$ and hence $S^+=\emptyset$ for $n\le 0$ and $S^-=\emptyset$ for $n\ge 0$.
Summing up, the following characterization of $S^\pm$
\begin{equation}\label{Spm}
 f\in S^\pm
 \qquad\text{if and only if}\qquad
 f\in \mathcal{L}_\R
 \big\{U_\nu,V_\nu\big\}_{-\frac 12\le \nu\le \pm\frac{1}{2}n-1},
\end{equation}
holds. In the last expression $\mathcal{L}_\R$ indicates the real linear combinations and
$$
 U_\nu =
 \begin{cases}
  H_\nu^{\text{o}}+H_{-1-\nu}^{\text{o}}
  &\text{if }r>r_1,
  \\
  H_\nu^{\text{i}}+H_{-1-\nu}^{\text{i}}
  &\text{if }r\le r_1
 \end{cases},
 \qquad
 V_\nu =
 \begin{cases}
  iH_\nu^{\text{o}}-iH_{-1-\nu}^{\text{o}}
  &\text{if }r>r_1,
  \\
  iH_\nu^{\text{i}}-iH_{-1-\nu}^{\text{i}}
  &\text{if }r\le r_1
 \end{cases}
$$
with $H_\nu^{\text{i}}= K_\nu(r_1) e^{i\nu\theta}(r/r_1)^{\nu+|n|/2}$, which is the holomorphic extension to the inner part of $e^{\pm i n\theta/2}H_\nu^{\text{o}}$.

By substituting in \eqref{zmspace}
and by defining $\mathcal{Z}(\mathbf{r})=(1-i) e^{i|n|\theta/2}f(hv\mathbf{r})$ with $f\in \{U_\nu,V_\nu\}$
the generators of the zero mode space
\begin{equation}\label{zmspaced}
\psi= \begin{pmatrix}
  \mathcal{Z}
  \\
  0
  \\
  0
  \\
  \mathcal{Z}^*
 \end{pmatrix}
 \ \text{ if }n<0,
 \quad
\psi= \begin{pmatrix}
  0
  \\
  -i\mathcal{Z}^*
  \\
  i\mathcal{Z}
  \\
  0
 \end{pmatrix}
 \ \text{ if }n>0,
 \quad\text{ for}\quad -\frac12\le \nu\le \frac{|n|-2}{2},
\end{equation}
are obtained. There are $|n|/2+\delta$ values of $\nu$ in the range
and the $U_\nu$ and $V_{\nu}$ form a linearly independent set except for $V_{-1/2}=0$ occurring when $n$ is odd. In total there are $2(|n|/2+\delta)-2\delta=|n|$ independent zero modes. Note that
the operator $i\sigma^1\otimes\sigma^3$ passes the zero modes for $-n$ to those for $n$.

Qualitatively the above results match the formal arguments given  in \cite{JaRo}, for instance about the number of zero modes and the structure of them. 
These arguments follow from the application of {the outstanding  index theorem. In that reference, the problem is addressed for $B(r)$ and $F(r)$ generic smooth functions. The explicit expressions of these zero modes
were found explicit due to our specialization to the step function model.  The advantage of the present complex variable approach is  that reduces the amount of calculations.

\subsection{Some examples of zero modes}

In order to make the statements given above less abstract, it may be of interest to derive some zero modes explicitly.
For $n$ odd the resulting zero modes are elementary functions  for~$\nu$ half-integral, as follows from  the formulas (8.468) and (8.486.16) of \cite{GrRy}
$$
 K_\nu(r)
 =
 \sqrt{\frac{\pi}{2r}}
 e^{-r}
 \sum_{m=0}^{|\nu|-1/2}
 \frac{(|\nu|+m-1/2)!}{m!(|\nu|-m-1/2)!}
 (2r)^{-m}.
$$
In particular, for any $n$,
$$
 \sqrt{\frac{2}{\pi}}
 H_\nu^{\text{o}}
 =
 \begin{cases}
  r^{-1/2} e^{-r\pm i\theta/2}
  &\text{if }\nu=\pm\frac 12,
  \\[5pt]
  r^{-3/2}(r+1)  e^{-r-3i\theta/2}
  &\text{if }\nu=-\frac 32
 \end{cases}
$$
and
$$
 \sqrt{\frac{2}{\pi}}
 H_\nu^{\text{i}}
 =
 \begin{cases}
  r_1^{-1/2}(r/r_1)^{(|n|\pm 1)/2} e^{-r_1\pm i\theta/2}
  &\text{if }\nu=\pm\frac 12,
  \\[5pt]
  r_1^{-|n|/2}(r_1+1) r^{(|n|-3)/2} e^{-r_1-3i\theta/2}
  &\text{if }\nu=-\frac 32.
 \end{cases}
$$
These expressions allows to work out fully the zero modes \eqref{zmspaced} for any $n$. Below, they will be specialized for $n=\pm1$ and $n=\pm3$.

For $n=\pm1$ the range for $\nu$ forces $\nu =-1/2$ and thus $V_{-1/2}=0$ and, according to the previous formulas,  $U_{-1/2}$ equals, except for a real constant factor,
$$
 f
 =
 \begin{cases}
  r^{-1/2} e^{-r-i\theta/2}
  &\text{if }r>r_1,
  \\
  r_1^{-1/2} e^{-r_1-i\theta/2}
  &\text{if }r\le r_1.
 \end{cases}
$$
The corresponding (not normalized) zero modes are in the outer part
\be\lb{zermod}
\Psi_0= \frac{e^{-hvr}}{\sqrt{hvr}}
 \begin{pmatrix}
  0
  \\
  1-i
  \\
  1+i
  \\
  0
 \end{pmatrix}
 \quad\text{for $n=1$},
 \qquad
 \Psi_0=\frac{e^{-hvr}}{\sqrt{hvr}}
 \begin{pmatrix}
  1-i
  \\
  0
  \\
  0
  \\
  1+i
 \end{pmatrix}
 \quad\text{for $n=-1$},
\ee
while in the inner part they are the constants corresponding to substitute $r=r_0$ (recall that $r_1=hvr_0$, then $f(hv\mathbf{r})$ changes its definition on $r=r_0$).
Note that, up to an irrelevant complex factor, the solution with $n=1$ coincides with the one found in the reference  \cite{zeromode}.
The resulting spinor components are purely radial in this case,  but the situation is different for higher values of $|n|$.
\smallskip

For $n=\pm3$, $\nu\in\{-1/2,1/2\}$. For $\nu=-1/2$ again $V_\nu=0$ and, omitting a real constant factor, $U_\nu$ is
$$
 f
 =
 \begin{cases}
  r^{-1/2} e^{-r-i\theta/2}
  &\text{if }r>r_1,
  \\
  r_1^{-3/2}r e^{-r_1-i\theta/2}
  &\text{if }r\le r_1.
 \end{cases}
$$
On the other hand, for $\nu=1/2$, $U_\nu$ is a real constant multiple of
$$
 g
 =
 \begin{cases}
  r^{-1/2} e^{-r+i\theta/2}
  +
  r^{-3/2}(r+1) e^{-r-3i\theta/2}
  &\text{if }r>r_1,
  \\
  r_1^{-5/2}r^2 e^{-r_1+i\theta/2}
  +
  r_1^{-3/2}(r_1+1) e^{-r_1-3i\theta/2}
  &\text{if }r\le r_1
 \end{cases}
$$
and $V_\nu$ is a real multiple of
$2^{-1/2}e^{\pi i/4} g(r,\theta+\pi/2)$
because
$e^{\pi i/4}e^{\pi i/4}$ and $e^{\pi i/4}e^{-3 \pi i/4}$ introduce the required $i$ and $-i$ factors.
Hence, the zero modes for $n=3$ are
$$
 \begin{pmatrix}
  0
  \\
  Z_1
  \\
  Z_1^*
  \\
  0
 \end{pmatrix}
  ,
 \
 \begin{pmatrix}
  0
  \\
  Z_2
  \\
  Z_2^*
  \\
  0
 \end{pmatrix}
  ,
 \
 \begin{pmatrix}
  0
  \\
  Z_3
  \\
  Z_3^*
  \\
  0
 \end{pmatrix}
  \quad\text{with }
 \begin{cases}
  Z_1= (1-i)e^{-3i\theta/2}f^*(hv\mathbf{r})
  \\
  Z_2= (1-i)e^{-3i\theta/2}g^*(hv\mathbf{r})
  \\
  Z_3= ie^{-3i\theta/2}g^*\big(hvr,\theta+\frac{\pi}{2}\big)
 \end{cases}
$$
For $n=-3$, there are similar expressions, which, according to \eqref{zmspaced}, are obtained applying to these spinors the operator
$-i\sigma^1\otimes\sigma^3$.

\subsection{On the number of bound states}
The zero modes have been characterized above. The remaining part of the bound state spectrum is difficult to be found explicitly.
However, some words can be said about the number of such eigenvalues.

From the energy equation \eqref{eneq} for $n=1$ it is possible to infer that for each $k$ fixed
\begin{equation}\label{asymp}
 \#\{
 \text{bounded states}
 \}
 \sim
 \frac{2hv}{\pi r_0}
 \qquad\text{for }
 hvr_0^{-1}\text{ large}
\end{equation}
and there are not bounded states, apart from the zero mode, for $hvr_0^{-1}$ small. In fact, the argument below suggests that the error term in this asymptotic formula is $O(1)$. This is related to a claim made in \cite[p.\,367]{rubakov} through dimensional analysis.
\medskip

Note that the number of solutions of \eqref{eneq} is invariant by the scaling $\lambda\mapsto \lambda r_0^{-1}$, $hv\mapsto hv r_0^{-1}$,
then it is enough to show \eqref{asymp} when $r_0=1$.
Another reduction is to assume $k\ge 0$ because \eqref{eneq} is invariant by $k\leftrightarrow -k$ \cite[8.404.2,8.486.16]{GrRy}. The case $k=0$ is somewhat special and it will treated at the end.

If $hv$ is small, using that
\[
 \frac{K_{k-1/2}(x)}{K_{k+1/2}(x)}
 \sim \frac{x}{k}
 \qquad\text{and}\qquad
 J_k(x)\sim
 \frac{x^k}{2^k k!}
 \qquad\text{when}\quad x\to 0,
\]
the energy equation (with $r_0=1$) translates asymptotically into $c_1\lambda^{2k-2}=c_2\lambda^{2k}$ for some positive constants $c_1$ and $c_2$ depending on $k$ and it has no solution for $\lambda>0$ small.

If $hv$ is large, typically, in the most of the range, $\lambda$ and $\lambda_{hv}=\sqrt{h^2v^2-\lambda^2}$ are large too. Let us now deal with this situation, which is the main case, and the rest of the cases will be considered later.
The asymptotics
\[
 \frac{K_{k-1/2}(x)}{K_{k+1/2}(x)}
 \sim 1
 \qquad\text{and}\qquad
 J_k(x)\sim
 \sqrt{\frac{2}{\pi x}}
 \cos\Big(x-\frac{\pi}{2}k-\frac{\pi}{4}\Big)
 \qquad\text{when}\quad x\to \infty,
\]
show
\[
 \Big(
 \frac{K_{k-1/2}(\lambda_{hv})}{K_{k+1/2}(\lambda_{hv})}
 J_{k-1}(\lambda)
 -
 \frac{K_{k+1/2}(\lambda_{hv})}{K_{k-1/2}(\lambda_{hv})}
 J_{k+1}(\lambda)
 \Big)
 J_k(\lambda)
 \sim
 \frac{2(-1)^k}{\pi\lambda}\cos(2\lambda)
\]
and
\[
 J_k^2(\lambda)+J_{k-1}(\lambda)J_{k+1}(\lambda)
 \sim
 \frac{2(-1)^k}{\pi\lambda}\sin(2\lambda).
\]
Substituting this in the energy equation \eqref{eneq}, it behaves asymptotically as
\[
 \frac{\lambda}{\sqrt{h^2v^2-\lambda^2}}
 =
 \tan(2\lambda).
\]
Now the argument is like in the standard treatment of the finite potential well appearing in basic quantum mechanics textbooks. The function $\tan(2\lambda)$ has $\frac{\pi}{2}$-spaced vertical asymptotes and each of them is cut by the graph of the function in the left hand side. So, the spacing of the eigenvalues is asymptotically $\pi/2$, showing \eqref{asymp}.

Still when $hv$ is large, there are two marginal situations to consider. We discuss them briefly. For small values of $\lambda$, again the first term in the Taylor expansion of the $J$-Bessel function shows as before that there are not solutions. It may also occur $\lambda$ large and $\lambda_{hv}$ small, if $\lambda$ is very close to $hv$. In this case, the first quotient of the $K$-Bessel functions in \eqref{eneq} is dominant because it behaves as $k/\lambda_{hv}$, but $\lambda_{hv}<c$ implies $\lambda =hv+O\big(h^{-1}v^{-1}\big)$, then the $J$-Bessel functions cannot complete full oscillations and the number of solutions is bounded.
\smallskip

Finally, the case $k=0$ is not covered by the previous treatment when $\lambda$ is small since $k$ denominators may appear. In this case, using $K_{-1/2}=K_{1/2}$ and $J_{-1}=J_{1}$, the energy equation becomes
\[
 \frac{2\sqrt{h^2v^2-\lambda^2}}{\lambda}
 =
 \frac{J_1(\lambda)}{J_0(\lambda)}
 -
 \frac{J_0(\lambda)}{J_1(\lambda)}
\]
and employing that the zeros of $J_0$ and $J_1$ are simple and asymptotically equal to
$\big(m-\frac 14\big)\pi$
and
$\big(m+\frac 14\big)\pi$
respectively \cite[8.547]{GrRy}, it is obtained again a collection of vertical asymptotes with a spacing tending to $\pi/2$.
\medskip

A final comment is that the employed estimates are not uniform in $k$ and hence \eqref{asymp} does not remain valid when $k$ grows. The functions $J_k(x)$ change dramatically its behavior when $k$  surpasses a short transition range  around $x$. It loses its oscillation decaying exponentially \cite[8.455.1]{GrRy} and then the solutions of the energy equation tend to disappear.
Taking this into account, $k$ is essentially limited to $hv$ and, if some uniformity can be established in this range,  the $\pi/2$-spacing detected by each $k$ should be replaced by an average spacing of order  $h^{-1}v^{-1}$ when $k$ is allowed to vary.

\section{The superconducting string absorption cross section for fermions}
After the previous section characterization of the energy levels and the zero modes, the next task is to study the absorption process of fermions by the vortex.
These quantities are important for this task.

Consider first a naked superconducting string, that is, a string which did not absorb yet any fermion. Such string has a set of energy levels $E_n=\sqrt{\lambda^2_n+p_z^2}$ inside its core, which are to be filled with fermions.
This would lead to a large current inside the object, specially from those fermions occupying the zero modes. As it will be clear from the discussion below, the Fermi line of zero modes has ``radius'' $E_f=2p_z=E_i$,
with $E_i$ the initial fermion energy. The value of this radius follows from the fact that the impulse in $p_z$ is not fixed, but only bounded by the energy conservation condition
in the absorption process $\psi_f+$vortex$\to\gamma+\psi_b$+vortex, with $\psi_f$ and $\psi_b$ free and bounded fermions states. In addition, $\gamma$ denotes an abelian gauge field excitation, which is not necessarily identified with an ordinary photon.
The current production will be effective  if the filling of levels is effective as well, and this will be the case  if the absorption scattering cross section is not suppressed. For this reason, specially in the
context of vorton formation, the study of this cross section may be relevant. This will be studied below for light fermions $m r_0<<1$ or heavy fermions $m r_0>>1$, and the intermediate case $m r_0\sim 1$. Some of the arguments of \cite{vorton3}-\cite{vorton5}
are related to the intermediate case, which is a situation more or less reasonable for solid state physics. However, there is no reason to confine the attention only to this case. For instance, a large coupling with the curvature of the space time $R$
may induce a large $mr_0$ value, or an effective suppressed coupling may induce a very small value of this parameter. For this reason, the discussion given. below will be general, not related to a particular $mr_0$ value.

\subsection{General formulas for the cross section}
The process to be considered in the following resembles partially the photoelectric effect, which corresponds to a  process in which  an atom captures a photon and emits an electron, which is initially in a bound state and 
then escapes to the continuous spectrum. In the present case the role of the atom is played by the vorton, but the effect to be considered is the inverse one, that is, the absorption of a fermion, which consequently falls into the bound state $\lambda<hv$. This process is accompanied by a photon or scalar boson emission. However, the former possibility is expected to be leading because massive particles are more difficult to excite. For this reason, the following discussion will be centered in the  photon emission. The differential cross section for this process can be worked out by analogy with the photoeffect, and the result is
\be\lb{fanky}
d\sigma=\frac{2\pi V}{v_e} | V_{fi}|^2\delta(E_\gamma+E_f-E_i) \frac{Vd^3k L_v dp_z}{(2\pi)^4}.
\ee
Here $\overline{k}$ denotes the photon impulse, $v_e$ is the initial electron velocity, $E_\gamma=\omega=|\overline{k}|$
is the photon energy $\omega=|\overline{k}|$ and
\be\lb{param}
E_i=\sqrt{p_x^2+m^2},\qquad E_f=\sqrt{\lambda_n^2+p_z^2},
\ee
are the initial and final fermion energies. The fermion impulse inside the vorton $p_z$ has to be integrated, since its possible values are bounded but continuous. On the other hand, $|V_{fi}|^2 \delta(E_\gamma+E_f-E_i)$ denotes the transition probability for unit time for the process. This matrix can be defined as follows.
First, the photon free field is described as
\be\lb{A}
\overline{A}=\frac{\overline{\epsilon}_i}{\sqrt{\omega V}} e^{i\overline{k}\cdot\overline{r}}
\ee
Here the polarization vectors $\epsilon_i$ with $i=1,2$ are orthogonal to the wave propagation direction $\overline{k}$ 
and orthogonal between them. In other words
$\overline{\epsilon}_i\cdot\overline{\epsilon}_j=\delta_{ij}$ and  $\overline{\epsilon}_i\cdot \overline{k}=0$. 
In addition, the incoming fermion field is free, that is, $\lambda>hv$ and, by assuming that it travels through the $x$-axis,
it is
given by a superposition of the following options
\be\lb{start}
\psi^\pm_s=u^\pm_s e^{-i p_x x},\qquad 
u^+_s=\frac{m}{\sqrt{2V}E_i}\left(
\begin{array}{c}
  1  \\
 0  \\
 \frac{E_i}{m}\\
\frac{p_{x}}{m} 
\end{array}\right),\qquad u^-_s=\frac{m}{\sqrt{2V}E_i}
\left(
\begin{array}{c}
  0  \\
 1  \\
\frac{p_x}{m}\\
\frac{E_i}{m}
\end{array}\right).
\ee
On the other hand, the final fermion state $\psi_f$ has $\lambda< hv$, and corresponds to a discrete energy state. More precisely, it is given by
\be\lb{confinado}
\psi_f=\frac{1}{\sqrt{L_v}}\psi^T_f(x,y) e^{-i p_z z},
\ee
with $\psi_f^T(x,y)$ denoting one of the solutions found in the previous sections, and  $L_v$ being the vortex length, which is very large. Note that the $T$ in $\psi_f^T$ does not denote a transpose.
The wave function in the discrete spectrum will be normalized by the condition
$\int_{\R^2} \psi^{T\dag}_f \psi^T_f dx dy=1$.
In these terms the transition matrix is defined by
$$
V_{fi}=\int_V \overline{\psi}_f \gamma^\mu\psi_s  A_\mu d^3x=\frac{2\pi}{\sqrt{L_v}} \delta(k_z-p_z)\int_{\R^2} \overline{\psi}^T_f \gamma^\mu \psi_s A^T_\mu \;dx dy,
$$
where the Fourier representation of the Dirac delta function was employed and
\be\lb{A2}
\overline{A}^T= \frac{\epsilon_i}{\sqrt{\omega V}} e^{-i k_x x-i k_y y},
\ee
denotes the vector potential without the term $e^{i k_z z}$. It is clear that $|V_{fi}|^2\sim \delta^2(k_z-p_z)/L_v=\delta(k_z-p_z)\delta(0)/L_v$. The quantity $\delta(0)$
is divergent and has length units, and cancel the factor $L_v$. The term $2\pi\delta(k_z-p_z)$ can be considered as a model for the integral of $e^{i(k_z-p_z)z}$ along the vortex, which is assumed to be very long, and then it makes sense to change its square, appearing in $|V_{fi}|^2$, by the integral of $L_ve^{i(k_z-p_z)z}$, which is $2\pi L_v\delta(k_z-p_z)$. 

The volume $V$ cancels in the final expression for the scattering cross section. The point of making explicit these factors is that the quantity $|V_{ij}|^2$ becomes proportional to the transition probability for unit time.
For simplicity,  the following formulas are adapted to the case $V=1$. This leads to the following expression for the differential cross section per unit of length $d\widetilde{\sigma}=L_v^{-1}d\sigma$ after cancelling all the volume factors
\be\lb{expreso}
d\widetilde{\sigma}=\frac{1}{v_e (2\pi)^2}\bigg|\int_{\R^2} \overline{\psi}^T_f \gamma^\mu u_s \epsilon_{i\mu} \: e^{ip_x x-i k_x x-i k_y y}\;dx dy\bigg|^2\delta(E_\gamma+E_f-E_i) \delta(k_z-p_z)\frac{d^3kdp_z}{\omega}.
\ee
The expression (\ref{expreso}) has to be summed over all the initial spins and all the polarizations. 
The sum over all the polarizations $\epsilon_i$ can be performed by use of the relation
$$
\sum_{i=1}^2 \epsilon_{ia}\epsilon_{ib}=\delta_{ab}-\frac{k_a k_b}{|\overline{k}|^2},
$$
valid in the transverse gauge employed here, with $a$ and $b$ spatial indices. 
The  integral term in (\ref{expreso}) is  related to the Fourier transform of the spinor components. Then, by summing over the final states $u_i^*$ and $u_i^-$
it is found after some calculations that
$$
2\sum_{j=1}^2\sum_{a=+}^-\bigg|\int_{\R^2} \overline{\psi}^T_f\gamma^\mu u^a_s \epsilon_{j \mu} \: e^{i(p_x - k_x)x-i k_y y}\;dx dy\bigg|^2=(|\chi^f_1|^2+|\chi^f_2|^2)\frac{2m^2}{E_i^2}
$$
$$
+|\eta^f_1|^2\bigg[1+\frac{p_x^2}{E_i^2}-\frac{2p_x k p_z\cos\alpha}{E_i|\overline{k}|^2}\bigg]+|\eta^f_2|^2\bigg[1+\frac{p_x^2}{E_i^2}+\frac{2p_x k p_z\cos\alpha}{E_i|\overline{k}|^2}\bigg]
$$
$$
-\eta_1^{f\ast}\eta^f_2\frac{p_x}{E_i}\bigg[\frac{p_x p_z k e^{-i\alpha}}{E_i|\overline{k}|^2}+\frac{k^2 e^{2i\alpha}}{|\overline{k}|^2}\bigg]
-\eta_1^{f}\eta^{f\ast}_2\frac{p_x}{E_i}\bigg[\frac{p_x p_z k e^{i\alpha}}{E_i|\overline{k}|^2}+\frac{k^2 e^{-2i\alpha}}{|\overline{k}|^2}\bigg]
$$
$$
-\chi_2^{f\ast}\eta^f_2\frac{2m}{E_i}\bigg[1-\frac{2m p_xp_z k e^{i\alpha}}{E_i|\overline{k}|^2}\bigg]-\chi_2^{f}\eta^{f\ast}_2\frac{2m}{E_i}\bigg[1-\frac{2m p_xp_zk e^{-i\alpha}}{E_i|\overline{k}|^2}\bigg]
$$
\be\lb{polaro}
+(\chi_2^{f\ast}\eta^f_1+\chi_1^{f\ast}\eta^f_2)\frac{m p_x}{E_i^2}\bigg[1+\frac{p_z^2}{|\overline{k}|^2}+\frac{k^2 e^{-i\alpha}}{|\overline{k}|^2}\bigg]+(\chi_2^{f}\eta^{f\ast}_1+\chi_1^{f}\eta^{f\ast}_2)\frac{mp_x}{E_i^2}\bigg[1+\frac{p_z^2}{|\overline{k}|^2}+\frac{k^2 e^{i\alpha}}{|\overline{k}|^2}\bigg].
\ee
Here the spinor coordinates $\chi_j$ and $\eta_j$ are the two dimensional transforms of the spinor component in the transferred momentum space. Namely, they are functions of $(p_x-k_x,p_y)$, the transferred momentum from the initial
fermion to the abelian gauge field. Note that  $k_z$ was replaced by $p_z$, due to the delta term $\delta(p_z-k_z)$ in (\ref{expreso}). In addition $k=\sqrt{k_x^2+k_y^2}$ is the absolute value of the photon impulse
projection over the plane $xy$ and the angle $\alpha$ is defined by $k_x=k\cos\alpha$ and $k_y=k\sin \alpha$.
Therefore the quantity of interest is given by
\be\lb{expreso22}
2d\widetilde{\sigma}=\frac{2}{v_e(2\pi)^2}\sum_{i=1}^2\sum_{a=+}^-\bigg|\int_{\R^2}\overline{\psi}^T_f \gamma^\mu u^a_s \epsilon_{i\mu} \: e^{ip_x x-i k_x x-i k_y y}\;dx dy\bigg|^2\delta(E_\gamma+E_f-E_i)\frac{dk_x dk_y dp_z}{|\overline{k}|},
\ee
together with (\ref{polaro}).

When a beam of fermions is incident over the vorton, some of them will be absorbed by the object and will start to fill all the energy levels.  The Pauli exclusion principle entails
that they may occupy a given level $\lambda_n$ if the value of the impulse or spin direction are different. The whole discrete spectrum is to be filled by fermions, and the ones occupying the zero mode will move at the light speed because $E_f=p_z$. In the following, the contribution of these modes will be taken as leading.

From now on attention will be focused on the case of winding number one. The corresponding zero mode \eqref{zermod} can be normalized as $\Psi_0^T= (0, -if, f, 0)$ with
\be\lb{zermodn}
f(r)=\frac{1}{\sqrt{\pi r_0 (r_0+\frac{1}{m})}},\qquad r<r_0, \qquad f(r)=\frac{e^{-m (r-r_0)}}{\sqrt{\pi r(r_0+\frac{1}{m})}}, \qquad r\geq r_0.
\ee
The differential cross section (\ref{expreso22}) simplifies considerably, due to the fact that two spinor component vanishes. 
The delta term $\delta(E_\gamma+E_f-E_i)$
implies that
$$
\sqrt{k^2+p_z^2}+|p_z|=E_i.
$$
This results in 
\be\lb{dif4}
k^2=E_i^2-2E_i |p_z|,\qquad k dk=-E_i d|p_z|\qquad |\overline{k}|=E_i-|p_z|.
\ee
The last relations imply that  $0<k<E_i$ and that $-E_i<2p_z<E_i$, and that the values of these variables are related to each other.
Therefore,  using the general change of variables formula $\delta\big(g(t)\big)=|g'(t)|^{-1}\delta\big(t-g^{-1}(0)\big)$ for the $\delta$ function
$$
\delta(E_\gamma+E_f-E_i)\frac{dk_x dk_y dp_z}{|\overline{k}|}=\frac{1}{E_i}\bigg[\delta\bigg(p_z-\frac{E_i^2-k^2}{2E_i}\bigg)-\delta\bigg(p_z+\frac{E_i^2-k^2}{2E_i}\bigg)\bigg]k dk dp_z d\alpha.
$$
In these terms the cross section by unit length is
$$
2d\widetilde{\sigma}=\frac{1}{ v_e(2\pi)^2 E_i}\bigg\{|\chi^f_2|^2\frac{2m^2}{E_i^2}
+|\eta^f_1|^2\bigg[1+\frac{p_x^2}{E_i^2}+\frac{k}{E_i}\bigg(\frac{E_i^2-k^2}{E_i^2+k^2}\bigg)^2\bigg]
$$
$$
+\chi_2^{f\ast}\eta^f_1\frac{m p_x}{E_i^2}\bigg[1+\bigg(\frac{E_i^2-k^2}{E_i^2+k^2}\bigg)^2+\frac{4E_i^2 k^2 e^{-i\alpha}}{(E_i^2+k^2)^2}\bigg]
$$
$$
+\chi_2^{f}\eta^{f\ast}_1\frac{m p_x}{E_i^2}\bigg[1+\bigg(\frac{E_i^2-k^2}{E_i^2+k^2}\bigg)^2+\frac{4 E_i^2k^2 e^{i\alpha}}{(E_i^2+k^2)^2}\bigg]\bigg\}k dk  d\alpha.
$$
A further simplification is obtained by taking into account that the zero mode is entirely described by the Fourier transform of the function $f(r)$ in (\ref{zermodn}).
For this mode, $\chi_2=-i\eta_1$, and the same relation holds for their Fourier transforms, which are given by the transform $F(p_x-k_x, p_y)$ of $f(r)$ up to an imaginary unit.
This in particular implies that the fourth and the sixth terms in the last expression cancel, as they are purely imaginary.
Thus, the simpler expression for the scattering cross section for unit length
\be\lb{ssimple}
2d\widetilde{\sigma}=\frac{|F|^2}{(2\pi)^2 p_x}\bigg\{\frac{2m^2}{E_i^2}+1+\frac{p_x^2}{E_i^2}+\frac{k}{E_i}\bigg(\frac{E_i^2-k^2}{E_i^2+k^2}\bigg)^2
+\frac{8m p_x k^2 \sin\alpha}{(E_i^2+k^2)^2}\bigg\}k dk  d\alpha,
\ee
is obtained. Here the fact that $v_e E_i=p_x$ was employed.

\subsection{The estimation of the cross section}

The task is now to estimate $\widetilde{\sigma}$ by integration of \eqref{ssimple}. We would like to emphasize that we have made a large calculation by use of special functions, which allowed us to make the desired
estimation as a function of the dimensionless parameters $p_x r_0$, $m r_0$ and $E_x r_0$. However, after doing all this large calculation, we have found a way to make a quick estimation which essentially captures
all the relevant behavior. For this reason, this last calculation is the one employed below. It has the disadvantage of being less sophisticated or specific, but it has the advantage of capturing the main behavior
with a relatively compact procedure.
 
In the following, it will be shown that $c_1p_x^{-1}<\widetilde{\sigma}< c_2p_x^{-1}$ for some absolute positive constants $c_1$ and $c_2$. Actually, in the next arguments  the possible values of the involved constants are not very large and they could be explicitly displayed.

Note that when integrating \eqref{ssimple}, the term involving $\sin\alpha$  vanishes because $k_x=k\cos\alpha$, appearing in $|F|^2$ is even in $\alpha$. The expression $(E_i^2-k^2)/(E_i^2+k^2)$ lies between~$0$ and~$1$, then by the mean value theorem for integrals it can be extracted as a bounded quantity from the integral. Finally, it is obvious that $1\le 2m^2/E_i^2+p_x^2/E_i^2\le 2$. Taking these remarks into account, by defining the radial function $h:\R^2\longrightarrow\R$ having
\begin{equation}\label{f2h}
 h(r)=r_0f(r_0r)
\end{equation}
as profile, after the change of variables $k_x \mapsto \chi/r_0$, $k_y \mapsto \eta/r_0$, the cross section per unit of length becomes
\begin{equation}\label{csf}
 \widetilde{\sigma}
 \sim
 c
 p_x^{-1}
 \iint_D
 \big|\widehat{h}(\chi,\eta)\big|^2\,d\chi d\eta,
 \ee
where $D$ denotes the disk $D=\big\{(\chi-p_xr_0)^2+\eta^2\le E_i^2r_0^2\big\}$.
More precisely, the cross section is given by the last expression with $2<c<4$. Here $\chi$ and $\eta$ are by their definition dimensionless real variables and $\hat{h}(\chi, \eta)$ denote the Fourier transform of the 
function $h(r)$ defined in (\ref{f2h}). The disk $D$ comes form the limitation $k^2\le E_i$, which follows from equation \eqref{dif4}.

The square of the component $f(r)$ is normalized in the plane $xy$, as seen from equation \eqref{zermodn}. Therefore $h$ is also normalized in \eqref{f2h}. The Parseval-Plancherel identity applied to $h$ reads $$1=\int_{\R^2}|h|^2=(2\pi)^{-2}\int_{\R^2}|\widehat{h}|^2.$$Hence
$\widetilde{\sigma}<c_2p_x^{-1}$. The remaining part of the calculation is to show that there exists $c_1$ such that
\begin{equation}\label{cslower}
 \widetilde{\sigma}>c_1p_x^{-1}.
 \ee
Note the disk $D$ defined below \eqref{csf} contain the origin $(\chi, \eta)=(0,0)$ and its radius is at least $M_0=mr_0$, which is the minimal energy the particle may have.
Consider a centered disk $D_0$ of radius $R_0=\epsilon M_0$ with $0<\epsilon<1$, its specific value is to be chosen later. A simple draw shows that both $D$ and $D_0$ overlap at least on the sector 
of the plane $(\chi, \eta)$ defined by $-\pi/3\le \alpha\le \pi/3$ and $r< R_0$.
This fact and the purely radial dependence of $h$ imply that
$$
 \iint_D
 \big|\widehat{h}(\chi,\eta)\big|^2\,d\chi d\eta
 \ge
 \iint_{D_0}
 \big|\widehat{h}(\chi, \eta)\big|^2\,d\chi d\eta
 \ge
 \frac{2\pi}{3}
 \int_0^{\epsilon M_0}
 R\big|\widehat{h}(R)\big|^2\, dR,
$$
with $R=\sqrt{\chi^2+\eta^2}$. As the function $h(r)$ depends only on the radius, its transform  $\widehat{h}$ may be written as a Hankel transform \cite{taylor} as follows
$$\widehat{h}(R)=2\pi \int_0^\infty
vh(v)J_0(Rv)\, dv.$$ By taking into account the definitions of $f(r)$ and $h(r)$ given in \eqref{zermodn} and  \eqref{f2h}, and that the derivative of $xJ_1(x)$ is $xJ_0(x)$ then, by separating the part $v\in [0,1]$,
it is found that
\begin{equation}\label{fth}
 C\widehat{h}(R)=
 \frac{J_1(R)}{R}
 +I
 \ee
 with
$$
 I
 =
 \int_1^\infty
 \sqrt{v} e^{-M_0(v-1)}J_0(rv)\, dv,
 \qquad
 C = \frac{1}{2}
 \sqrt{\frac{M_0+1}{\pi M_0}}.
$$
For the case $M_0\le1$, it is seen after a change of variables that 
$$
 I =
 M_0^{-3/2}e^{M_0}
 \int_{M_0}^\infty
 \sqrt{v} e^{-v}J_0\Big(\frac{r}{M_0} v\Big)\, dv.
$$
This integral is estimated as  $$I =
 M_0^{-3/2}e^{M_0}\bigg[\int_{M_0}^\infty \sqrt{v} e^{-v}\, dv+O(\epsilon^2)\bigg],$$
for $0\le r\le \epsilon M_0$. This result can be found by taking into account that
$$J_0(x)= 1+O(x^2),\qquad J_1(x)/x\sim \frac 12+O(x^2),$$ the last  follows by employing that $J_1(x)=\frac 12 x+O(x^3)$.
By introducing this is \eqref{fth} and by taking into account that $C<M_0^{-1/2}$, the following estimation
$$
 \widehat{h}(R)>M_0^{-1}\int_1^\infty
 \sqrt{v} e^{-v}\, dv+O(\epsilon^2)
 >\frac 12 M_0^{-1},
 $$
 valid for $\epsilon$ small enough, is found. Therefore $\int_0^{\epsilon M_0}  R\big|\widehat{h}(R)\big|^2\, dR$ is greater than a positive constant, with values not far from unity,  and \eqref{cslower} is proved.

If instead $M_0\ge 1$ then by employing the same expansions of $J_0$ and $J_1$ as above, it is found for $0\le r\le\epsilon$ that $$I=\int_1^\infty \sqrt{v}e^{-M_0(v-1)}\, dv+O(\epsilon^2).$$ As $C\le 1/\sqrt{2\pi}$,
it is deduced that
$$
 \widehat{h}(R)>
 \frac 12 \sqrt{2\pi}+O(\epsilon^2)
 >1
 $$
 for $\epsilon$ small enough. Therefore $\int_0^{\epsilon M_0}  R\big|\widehat{h}(R)\big|^2\, dR\ge \int_0^{\epsilon }  R\big|\widehat{h}(R)\big|^2\, dR$ is greater than a positive constant, of the order of unity as well. 
 
 Overall, the above discussion shows that the cross section is given by $$\sigma=\frac{f(mr_0, p_x r_0)}{p_x},$$ where the function $f(m r_0, p_x r_0)$ parameterize the dependence on the radius 
 of the object. This function may be difficult to be found explicitly.  However, this function is bounded by two constants $c_1$ and $c_2$ which have moderate values, not very far from unity. Thus the cross section is 
 not very sensitive to the values of $r_0$, and it is divergent just as $p_x\to 0$ regardless the value of the mass $m$ or the radius $r_0$.

\section{The Aharanov-Bohm scattering cross section}
Having estimated the absorption cross section for fermions incident over the vorton, the next task is to compare it with the Aharonov-Bohm scattering cross section.
The exact form of this section is known for a delta type of singularity in $r$ \cite{alford}. As the interest of the present work is the behavior
of this section in several regimes involving $m r_0$ or $p_x r_0$, even taking into account that there exists literature about the subject,
it is convenient to make an independent estimation.

From the definition of the scattering section, an incident beam of fermions with energy $E>hv$ and momentum $p_x$ is dispersed by the vortex.
The calculation of this cross section entails the partial expansion of the wave function (\ref{start}), which is given by
$$
\psi^\pm_s=\left(
\begin{array}{c}
  \chi^\pm_1  \\
\chi^\pm_2  \\
\eta^\pm_1\\
\eta^\pm_2
\end{array}\right)=u_s^\pm e^{i p_x r\cos\theta}=u_s^\pm \sum_{n=-\infty}^\infty e^{\frac{in\pi}{2}} J_n(p_x r)  e^{in\theta}.
$$
For large $r$ values the asymptotic of the Bessel function $J_n(p_x r)$ shows that
\be\lb{asintop}
\psi^\pm_s\sim \frac{1}{2}u_s^\pm\sqrt{\frac{2}{\pi p_x r}} \sum_{n=-\infty}^\infty e^{\frac{in\pi}{2}} (e^{ip_x r-\frac{i\pi}{4}-\frac{in\pi}{2}}+e^{-ip_x r+\frac{i\pi}{4}+\frac{in\pi}{2}})  e^{in\theta}.
\ee
The last formula expresses the asymptotic wave function as a combination of an incident wave in $r$, proportional to $e^{-ip_x r}$, and a reflected one proportional to $e^{i p_x r}$.
The calculation of the cross section requires to find a solution of the equation of motion that asymptotically tends to this ingoing wave for large $r$. The outgoing wave
will be of course different than above, and this difference term is what determines the scattering cross section.

According to the appendix below, the component of the unbounded states admit linear formulas as expansions of the form
\[
 \sum_{n=-\infty}^\infty
 e^{in\theta}
 \Big(
 A_nJ_{n+\frac 12}\big(r\sqrt{\lambda^2-h^2v^2}\big)
 +
 B_nY_{n+\frac 12}\big(r\sqrt{\lambda^2-h^2v^2}\big)
 \Big)
\]
affected by constant factors and $e^{i\theta}$. Comparing the asymptotics of this to \eqref{asintop} as $r\to\infty$, it is deduced $p_x=\sqrt{\lambda^2-h^2v^2}$.
Then the calculations indicated in the appendix give
$$
\chi_1=\frac{e^{-i\theta}}{2}\sum_{n=-\infty}^\infty e^{i n\theta} \big[(d_{1n}+c_{1n}) J_{n-\frac12}(p_xr)+(d_{2n}+c_{2n}) Y_{n-\frac12}(p_xr)\big],
$$
$$
\chi_2=\frac{i}{2p_x}\sum_{n=-\infty}^\infty e^{i n\theta} \bigg[\big((E+m)d_{1n}+(E-m)c_{1n}\big) J_{n+\frac12}(p_xr)
+\big((E+m)d_{2n}+(E-m)c_{2n}\big) Y_{n+\frac12}(p_xr)\bigg],
$$
$$
\eta_1=\frac{1}{2}\sum_{n=-\infty}^\infty e^{i n \theta} \big[(d_{1n}-c_{1n}) J_{n-\frac12}(p_xr)+(d_{2n}-c_{2n}) Y_{n-\frac12}(p_xr)\big],
$$
\be\lb{abovo}
\eta_2=\frac{ie^{i\theta}}{2p_x}\sum_{n=-\infty}^\infty e^{i n \theta} \bigg[\big((E+m)d_{1n}-(E-m)c_{1n}\big) J_{n+\frac12}(p_x\;r)
+\big((E+m)d_{2n}-(E-m)c_{2n}\big) Y_{n+\frac12}(p_xr)\bigg],
\ee}
where $E_i$ is the initial particle energy and, as before, the volume will be taken as $V=1$.
A hint that the above formula is correct is related to the counting of unknowns, and goes as follows. The asymptotic behavior of the Bessel functions allows to fix the coefficients $c_{in}$ and $d_{in}$ with $i=1,2$ in order to match the behavior (\ref{asintop}) for the incoming wave. This in principle will give four equations. In addition, the inner part of the spinor components has in principle four unknowns $a_{in}$ and $b_{in}$ with $i=1,2$. These may be determined by the four continuity conditions at $r=r_0$. The apparent problem is that
the wave function is highly divergent at $r=0$ if the coefficients multiplying the Bessel function $Y_n(x)$ do not vanish. Thus, the situation is that there are four equations for two unknowns, and this may be problematic at first sight.
However, this is not the case, and this gives confidence about the correctness of (\ref{abovo}).
To see this, consider for instance the solution $\psi^+_s$. In this case the requirement that the incoming wave part in asymptotic form $e^{-i p_x r}$ has the form (\ref{asintop}) gives four equations for the four unknowns $c_{1n}$, $c_{2n}$, $d_{1n}$ and $d_{2n}$. However, the reader is challenged to check that only two equations are linearly independent, the other two do not give further information. 
The result is 
\be\lb{tuco}
c_{1n}+i c_{2n}=-\frac{E+m}{2\sqrt{2}E} e^{\frac{(2n+1) i \pi}{4}},\qquad d_{1n}+i d_{2n}=\frac{E-m}{2\sqrt{2}E} e^{\frac{(2n+1) i \pi}{4}}.
\ee
This means that two of these four coefficients are still undetermined. The introduction of the last formula into (\ref{abovo}) and the use of the asymptotic expressions for the Bessel functions shows that the incoming wave part is identical to the one in (\ref{asintop}).

On the other hand, the continuity condition at the border of the object, which is the region $r=r_0$, constitutes a non homogeneous system for $c_{1n}$, $c_{2n}$, $d_{1n}$ and $d_{2n}$,  and the constants $a_n$ and $b_n$ defining the core wave functions (\ref{generation1})-(\ref{generation2}) in the appendix. This, together with the last expression (\ref{tuco}), constitutes a 6 dimensional non homogeneous system with non trivial solution.  Before going about the explicit calculation of these coefficients, it may be convenient to characterize the scattering cross section for the process.
Consider the function
\be\lb{werden}
Z_{\nu}(x)=J_{\nu}(x)+iY_\nu (x)\sim\sqrt{\frac{2}{\pi x}} e^{ix-\frac{i\pi}{4}-\frac{\nu\pi}{2}}.
\ee
The relation (\ref{tuco}) allows to write
the wave function as
$$
\chi_1=\frac{e^{-i\theta}}{2}\sum_{n=-\infty}^\infty e^{i n\theta} \bigg[-\frac{m}{\sqrt{2}E}e^{(\frac{2n+1}{4})i\pi} J_{n-\frac12}(p_xr)
-i(d_{2n}+c_{2n}) Z_{n-\frac12}(p_xr)\bigg],
$$
$$
\chi_2=\frac{1}{2p_x}\sum_{n=-\infty}^\infty e^{i n\theta}\big((E+m)d_{2n}+(E-m)c_{2n}\big) Z_{n+\frac12}(p_xr)\bigg],
$$
$$
\eta_1=\frac{1}{2}\sum_{n=-\infty}^\infty e^{i n\theta} \bigg[\frac{1}{\sqrt{2}}e^{(\frac{2n+1}{4})i\pi} J_{n-\frac12}(p_xr)
-i(d_{2n}-c_{2n}) Z_{n-\frac12}(p_xr)\bigg],
$$
$$
\eta_2=\frac{ie^{i\theta}}{2p_x}\sum_{n=-\infty}^\infty e^{i n \theta} \bigg[\frac{p_x^2}{\sqrt{2}E}e^{(\frac{2n+1}{4})i\pi} J_{n+\frac12}(p_xr)
-i\big((E+m)d_{2n}-(E-m)c_{2n}\big) Z_{n+\frac12}(p_xr)\bigg].
$$
The asymptotics of $Z_\nu$ shows that the terms containing $Z_{n\pm 1/2}$, which are affected by the coefficients $d_{2n}$ and $c_{2n}$,
are outgoing waves. The remaining terms compose a mixture of ingoing and outgoing waves.
Thus, the determination of the coefficients $d_{2n}$ and $c_{2n}$ is fundamental for determining the scattering cross section.
This is given by \cite{alford}
$$
\frac{d\widetilde{\sigma}}{d\alpha}=\lim_{r\to \infty} \frac{r\cdot J_o}{-J_{ix}}, \qquad J_0=\overline{\psi}_o\gamma^\mu \psi_o,
$$
where $\psi_o$ is the outgoing contribution and $J_{ix}$ in the ingoing contribution. The result is
$$
\frac{d\widetilde{\sigma}}{d\alpha}=\lim_{r\to \infty}\frac{m (\eta_1 \eta_2^\ast- \chi_1 \chi^\ast_2)}{p_x}e^{i\theta}+h.c.
$$
The asymptotic form of the outgoing wave function that follows from (\ref{werden}) is
$$
\chi_{o1}\sim-\frac{1}{2}\sqrt{\frac{2}{\pi p_x r}}\sum_{n=-\infty}^\infty e^{i (n-1)\theta} \bigg[\frac{m}{2\sqrt{2}E} 
+(d_{2n}+c_{2n}) e^{-(\frac{2n-1}{4})i\pi}\bigg]e^{i p_x r+\frac{i\pi}{4}},
$$
$$
\chi_{o2}\sim-\frac{i}{2}\sqrt{\frac{2}{\pi p_x r}}\sum_{n=-\infty}^\infty e^{i n\theta}\bigg(\sqrt{\frac{E+m}{E-m}}d_{2n}+\sqrt{\frac{E-m}{E+m}}c_{2n}\bigg)e^{-(\frac{2n-1}{4})i\pi} e^{ip_x r+\frac{i\pi}{4}},
$$
$$
\eta_{o1}\sim\frac{1}{2}\sqrt{\frac{2}{\pi p_x r}}\sum_{n=-\infty}^\infty e^{i n\theta} \bigg[\frac{1}{2\sqrt{2}}
-(d_{2n}-c_{2n}) e^{-(\frac{2n-1}{4})i\pi}\bigg] e^{i p_x r+\frac{i\pi}{4}},
$$
\be\lb{nasim}
\eta_{o2}\sim\frac{i}{2}\sqrt{\frac{2}{\pi p_x r}}\sum_{n=-\infty}^\infty e^{i (n+1) \theta} \bigg[\frac{p_x}{2\sqrt{2}E}
-\bigg(\sqrt{\frac{E+m}{E-m}}d_{2n}-\sqrt{\frac{E-m}{E+m}}c_{2n}\bigg) e^{-(\frac{2n+1}{4})i\pi}\bigg]e^{i p_x r+\frac{i\pi}{4}}.
\ee
This asymptotic form will be fully characterized only by determining $c_{2n}$ and $d_{2n}$. This is achieved by use the continuity conditions at the border of the object.
These equations are given by
$$
a_n J_n(E r_0)-b_{n+1} J_{n+1}(Er_0)=\sqrt{\frac{E-m}{E+m}}\bigg[c_{1n} J_{\frac{2n+1}{2}}(p_x r_0)+c_{2n} Y_{\frac{2n+1}{2}}(p_xr_0)\bigg],
$$
$$
a_n J_n(E r_0)+b_{n+1} J_{n+1}(Er_0)=\sqrt{\frac{E+m}{E-m}}\bigg[d_{1n} J_{\frac{2n+1}{2}}(p_x r_0)+d_{2n} Y_{\frac{2n+1}{2}}(p_xr_0)\bigg],
$$
$$
a_n J_{n-1}(E r_0)-b_{n+1} J_{n}(Er_0)=c_{1n} J_{\frac{2n-1}{2}}(p_x r_0)+c_{2n} Y_{\frac{2n-1}{2}}(p_xr_0),
$$
\be\lb{conti}
a_n J_{n-1}(E r_0)+b_{n+1} J_{n}(Er_0)=d_{1n} J_{\frac{2n-1}{2}}(p_x r_0)+d_{2n} Y_{\frac{2n-1}{2}}(p_xr_0).
\ee
where the coefficients $a_n$ and $b_n$ correspond to the inner part as shown in the appendix and $\lambda$ has been identified with the energy $E$.
This calculation in the generic situation is not very illuminating, for this reason it is more convenient to consider it in several regimes, in which the Bessel functions
have limits which are expressed in terms of elementary functions. Consider first large values of $p_x$. By eliminating $a_n$ and $b_{n+1}$ respectively, the system reduces to
the two dimensional one
$$
-\tan(E r_0-\frac{\pi}{4}-\frac{n\pi}{2})\bigg[\bigg(\sqrt{\frac{E+m}{E-m}} d_{1n}+\sqrt{\frac{E-m}{E+m}} c_{1n}\bigg)\cos\alpha
+\bigg(\sqrt{\frac{E+m}{E-m}} d_{2n}+\sqrt{\frac{E-m}{E+m}} c_{2n}\bigg)\sin\alpha\bigg]
$$
$$
=(d_{1n}+c_{1n})\cos\alpha+(d_{2n}+c_{2n})\sin\alpha,
$$
$$
\cot(E r_0-\frac{\pi}{4}-\frac{n\pi}{2})\bigg[\bigg(\sqrt{\frac{E+m}{E-m}} d_{1n}-\sqrt{\frac{E-m}{E+m}} c_{1n}\bigg)\cos\alpha
+\bigg(\sqrt{\frac{E+m}{E-m}} d_{2n}-\sqrt{\frac{E-m}{E+m}} c_{2n}\bigg)\sin\alpha\bigg]
$$
$$
=(d_{1n}-c_{1n})\cos\alpha+(d_{2n}-c_{2n})\sin\alpha.
$$
where
\[
 \alpha= p_x r_0-\frac{\pi}{4}-\frac{(2n+1)}{4}\pi.
\]
The last two equations together with (\ref{tuco}) conform a non homogeneous four dimensional linear system for the four unknowns $d_{in}$ and $c_{in}$ with $i=1,2$.
The functional form of these coefficients can be found explicit, and it  is of course cumbersome. This complicates the hope of computing the sums in (\ref{nasim}. However, something interesting happens when
$$
p_x r_0=\frac{M\pi}{2},
$$
where $M$ is an integer. In this case, depending on the value of  $n$, either the sine or the cosine in the last expression vanishes and the result is an homogeneous system for either $c_{1n}$ and $d_{1n}$
or $c_{2n}$ and $d_{2n}$. For a generic energy, either one pair of coefficients or the other vanish. The non vanishing pair is then found by use of (\ref{tuco}). This means that the coefficients for $n=2m$
or $n=2m+1$ have the same functional form, except for a multiplication by $-i$.  Therefore, for the discrete set of impulses corresponding to this situation, the asymptotic form  (\ref{nasim}) is simply
 $$
\chi_{o1}=-\frac{1}{2}\sqrt{\frac{2}{\pi p_x r}}\frac{me^{i p_x r+\frac{i\pi}{4}}}{2\sqrt{2}E}(1-i e^{i\theta}) \sum_{m=-\infty}^\infty e^{i (2m-1)\theta},\qquad 
\chi_{o2}=-0,
$$
\be\lb{nasim2}
\eta_{o1}=\frac{1}{2}\sqrt{\frac{2}{\pi p_x r}}\frac{e^{i p_x r+\frac{i\pi}{4}}}{2\sqrt{2}} (1-i e^{i\theta}) \sum_{m=-\infty}^\infty e^{2im\theta},\qquad
\eta_{o2}=\frac{i}{2}\sqrt{\frac{2}{\pi p_x r}}\frac{p_x e^{i p_x r+\frac{i\pi}{4}}}{2\sqrt{2}E}(1-i e^{i\theta})  \sum_{m=-\infty}^\infty e^{(2m+1)i \theta}.
\ee
The angular sums are divergent, proportional to the Dirac delta $\delta(\theta)+\delta(\theta-\pi)$. This is expected since it is known that the Aharonov-Bohm total scattering section possesses this behavior \cite{ruijsenaars}.
The cross section can be renormalized as
\be\lb{biode}
\frac{d\widetilde{\sigma}}{d\theta}\sim\frac{m}{2E p_x}\frac{\sin\theta(1-\sin\theta) \sin^2(2L+1)\theta}{\sin^2\theta}, \qquad L>>1,
\ee
with $L$ a cutoff, which has to be sent to infinite when integrating over $\theta$. The proportionality factor to the angular divergent term goes as
$$
\frac{m}{Ep_x}=
\frac{m}{p_x \sqrt{m^2+p_x^2}},
$$
which shows that if the particle is light and has high energy due to a large momentum, then this coefficient is suppressed. On the other hand, if the mass is very large, the factor is still suppressed by $p_x$, which means that for large momenta the cross section is suppressed regardless the particle is light or heavy. This conclusion only holds for the discrete set described above, nevertheless we expect the same behavior in the more generic situation, since we do not expect that the cross section will have enormous peacks in the middle value energies between this set.

The formula (\ref{biode}) is also approximately true for small $p_x$ when $m$ is large. The reason is that if $p_x r_0$ takes a small value, then (\ref{conti}) contains the highly divergent $Y_{\nu}(x)$ functions at the origin, which implies
that $c_{2n}$ and $d_{2n}$ are approximately zero. Thus, the asymptotic form (\ref{nasim2}) is still valid, up to a factor $(1-e^{i\theta})$ and (\ref{biode}) is deduced, up to the factor $(1-\sin \theta)$. For very small masses and impulses, the standard Aharonov-Bohm scattering cross section is expected. In this case the cross section goes as $1/p_x$. 

Based on this facts, it can be suggested that the differential cross section has the form
\be\lb{biode2}
\frac{d\widetilde{\sigma}}{d\theta}\sim\frac{1}{p_x}f(m, p_x) a(\theta), 
\ee
with $a(\theta)$ an angular factor whose integration over a period is divergent, and with a dimensionless function $f(x, y)$ such that  $f(0, 0)=1$ and $f(m, p_x)\sim \frac{m}{E_i}$ for $p_x$ large. 

\section{Discussion}
In the present work, the Dirac equation and the bound and scattering states for a vortex with winding number one was studied. In addition, the $n$ zero modes corresponding to the generic winding number $n$, under the crude assumption that the vortex profile is given by a step function, were constructed explicitly. Furthermore, the absorption cross sections for fermions incident on a vortex with generic
masses and impulses was estimated in several regimes.
The main formula is that the cross section for unit length is $\widetilde{\sigma}\sim f(mr_0, p_x r_0)/p_x$ with the function $f(m r_0, p_x r_0)$ taking moderate values, not far from unity, regardless the value of the parameters $p_x$, $m$ and $r_0$.
This means that if the object has a perimeter of a centimeter, then only if $p_x> M_p$ the cross section will be smaller than a typical nuclear one $\sigma_n \sim 10^{-26}$cm$^2$. Thus, the interaction of the fermions with these objects is noticeable.
However, if a nuclear cross section by unit length for nuclear forces $\sigma_n/r_n\sim 10^{13}$ cm, then an impulse $p_x\sim 10^{9}$GeV is enough for having a smaller value of the cross section by unit length.
Thus the interaction between fermions and these objects may depend on the density of this objects in comparison with other particles in the plasma which interact with those fermions. However, unless the plasma density is too high,
the absorption process will be effective and large currents will be induced on the object.

Given the cross sections found above, it may be of interest to discuss the force the vortex resulting from the absorption of fermions with gauge field emission, and to compare it to the one due to Aharonov-Bohm scattering.
It may be convenient to make a simple estimation instead of a lengthy specific calculation. First, as follows from \cite{pelea1}, \cite{kopnin19}, for Aharonov-Bohm scattering the transversal and longitudinal forces are determined by the differential scattering cross sections
$$
\frac{d\sigma_l}{d\alpha}= \frac{d\sigma}{d\alpha}(1-\cos\theta),\qquad \frac{d\sigma_t}{d\alpha}= \frac{d\sigma}{d\alpha}\sin\theta.
$$
Note that, even taking into account that the cross section $\sigma$ for the Aharonov-Bohm scattering (\ref{biode}), the sections $\sigma_t$ and $\sigma_l$ are convergent.
These sections of course, have to be multiplied by momentum factors and integrated along their possible values. They have also to be multiplied by factors describing the fermion density, since the larger the density is, the stronger the force will be over the vortex.
In fact, the formulas derived in those references are designed in order to match with the average of the momentum flux tensor $\Pi_{ij}=P\delta_{ij}+\rho v_i v_j$, with $P$ the plasma pressure, $v_i$ the fermion density and $\rho$ its mass density.
This flux has to be integrated over a surface surrounding the object $d\hat{S}_{i}$.

Based on the reading of those references, the estimation to be proposed is the following. It may be assumed that the photon mean value momentum  by unit time in some direction, say $\hat{x}$, is given by
$$
\frac{<k_x>}{T}=\frac{1}{\int \frac{|w(k)|^2}{T} d^2k}\int \frac{|w(k)|^2}{T} k_x d^2k
$$
where $|w(k)|^2/T$ is the transition probability for unit time.
The denominator in the last expression is due to the fact that it is not normalized, since there are other possible transitions than absorption.
An analogous formula holds for $k_y$.
In addition, the density of photons can be estimated as
$$
N_\gamma=N_f\int \frac{|w(k)|^2}{T} d^2k,
$$
with $N_f$ the fermion density.  Therefore one may introduce a quantity with force density units 
$$
<F_x>=\frac{N_\gamma <k_x>}{T}=\int \frac{|w(k)|^2}{T} k_x d^2k.
$$
 A similar quantity can be defined for the Aharonov-Bohm effect. On the other hand, the probability $|w(k)|^2/T$ is proportional to $|V_{ij}|^2\delta(E_i-E_f-k)$, and up to flux factors, 
 to the full scattering cross section (\ref{fanky}). Therefore, in order to compare the forces for the absorption process and the Aharonov-Bohm one along the direction $\hat{x}$, it
 is convenient to study the quotient
 $$
 \frac{F_{ax}}{F_{sx}}=\frac{\int (p-k\cos\alpha) d\sigma_a}{\int (1-\cos\alpha)p d\sigma_{s}},
 $$
 where the index $a$ denotes the absorption quantities and $s$ the scattering (Aharonov-Bohm) ones found along the text.
In fact, the transferred momentum is related to averages of $p-k\cos\alpha$ for the absorption process or to $p(1-\cos\theta)$ for scattering,
while the transversal is proportional to $k \sin\alpha$ or $p\sin\theta$.  Based on this, we made a calculation similar to those in the previous
 section, but with these factors included. 
We have considered only the longitudinal force, since the Aharonov-Bohm section is divergent but the multiplication by $(1-\cos\theta)$ cures the divergence.
For the transversal forces, one may assume that the magnitude orders are conserved, although a more careful analysis may be of interest. 
In any case, the repetition of all the arguments of the text with these factors now included leads us to the following upper bound
$$
\frac{F_{ax}}{F_{sx}}\geq \frac{c_1}{ f(m, p_x)}\bigg\{
\frac{4m^2 E_i^2-2p_x E_i (10 m^2-2p_x^2)+(m^2+p_x^2)(8m^2-4p_x^2)}{4m^2p_x\sqrt{m^2+(E_i-p_x)^2}}
$$
$$
-\frac{4m^2 p_x^2-2p^2_x (10 m^2-2p_x^2)+(m^2+p_x^2)(8m^2-4p_x^2)}{4p_xm^3}+3\log \bigg(\frac{2\sqrt{m^2+(E_i-p_x)^2}+2E_i-2p_x}{2m}\bigg)\bigg\}.
$$
$$
+\frac{c_2}{f(m, p_x)}\bigg[\frac{2m^2}{E_i^2}+1+\frac{p_x^2}{E_i^2}+\mu\bigg] \bigg[\pi-\sqrt{2}\sqrt{1+\frac{p_x}{E_i}} \; E\bigg(\frac{2 p_x}{E_i+p_x}\bigg)\bigg],
$$
valid for small or intermediate values $m r_0\leq 1$. The function $f(x,y)$ is the one defined in 
the Aharonov-Bohm cross section (\ref{biode2}) and $c_1$ and $c_2$ are constants, with moderate values. In addition, $E(k)$ is the second elliptic integral. It is seen that for large $p_x$, as $f\to m/E_i$, the quotient is very large. The particle with large $p_x$ is difficult to be absorbed
but once it is, it gives a large momentum to the vortex. For this reason, it becomes important. For small values of $p_x$, the absorption force seems also relevant. However, for this situation, the expressions that have been found above for the Aharonov-Bohm factor $f(m, E_i)$ are less trustable. 

For large masses $m r_0>>1$ the approximation goes as 
$$
\frac{F_{ax}}{F_{sx}}\geq \frac{c_1 E_i}{p_x}+c_2
\bigg[\frac{2m^2}{E_i^2}+1+\frac{p_x^2}{E_i^2}+\mu\bigg].
$$
This quotient takes considerable values when $p_x$ is very small, or when $p_x $ is very large. The physical interpretation for that is that for $p_x$ small, the particles which are very massive tend to be absorbed since
in this region they change abruptly the mass to a zero value, with a large gap, and are approaching slowly. The absorption in this case is favored and thus large forces on the vortex emerge. On the other hand, if $p_x r_0$ is larger than $m r_0$
then the absorption is suppressed but a single absorption involves a huge amount of energy which generates a large recoil in the vortex. 

It should be emphasized that, even taking into account that the references  \cite{kopnin0}-\cite{experimento} are part of the inspiration for the calculation of the above forces, the situation these works analyze is not exactly the same as here.
In those works, a beam of fermions is incident over a vortex with filled bound states, and the spectrum of bound states hypothetically has influence on the macroscopic force. In the present case, the vortex is initially naked, that is, it does not confine fermions. The vortex starts to absorb the fermions, which consequently fall into the bound spectrum, and this generates a recoil which is manifested in a macroscopic force. The possible implications of these results are model dependent, and are of interest for  a future work.

\section*{Acknowledgments}
O.P. S is grateful to the Universidad Autónoma and to the ICMAT de Madrid, where this work was performed, by their hospitality. The present work is supported by CONICET, Argentina and by the Grant PICT 2020-02181. This project has received funding from the European Union’s Horizon 2020 research and innovation program under the Marie Sk{\l}odowska-Curie grant agreement No 777822. F. Ch. is partially supported by the PID2020-113350GB-I00 grant of the MICINN (Spain) and
by ``Severo Ochoa Programme for Centres of Excellence in R{\&}D'' (CEX2019-000904-S).

\appendix

\section{The unbound states for $n=1$}
For winding number $n=1$, the  Dirac eigenfunction equation 
$\mathcal{D}\Psi=\lambda\Psi$ considered in the text with $\mathcal{D}$ defined in (\ref{main_eq})
can be given explicitly in polar coordinates as
$$
\bigg[-i e^{-i\theta} \partial_r-\frac{e^{-i\theta}}{r}\partial_\theta-\frac{ie^{-i\theta} B(r)}{2r}\bigg]\chi_2-hv e^{-i\theta}F(r)\eta_1=\lambda \chi_1,
$$
$$
\bigg[-ie^{i\theta} \partial_r+\frac{e^{i\theta}}{r}\partial_\theta+\frac{ie^{i\theta} B(r)}{2r}\bigg]\chi_1-hv e^{-i\theta}F(r)\eta_2=\lambda \chi_2,
$$
$$
\bigg[ie^{-i\theta} \partial_r+\frac{e^{-i\theta}}{r}\partial_\theta-\frac{ie^{-i\theta} B(r)}{2r}\bigg]\eta_2-hv e^{i\theta}F(r)\chi_1=\lambda \eta_1,
$$
\be\lb{sistem}
\bigg[ie^{i\theta} \partial_r-\frac{e^{i\theta}}{r}\partial_\theta+\frac{ie^{i\theta} B(r)}{2r}\bigg]\eta_1-hv e^{i\theta}F(r)\chi_2=\lambda \eta_2.
\ee
Consider the solution outside first. First, note that if the first two equations  (\ref{sistem}) are multiplied by $i$ and the last two ones by $-i$, and the specific profiles $B=F=1$ are chosen,
then the identities $e^{i\theta}\partial_\theta f=-i f+\partial_\theta (e^{i\theta} f)$ and $e^{-i\theta}\partial_\theta f=i f+\partial_\theta (e^{-i\theta} f)$, when applied in the second and third equation respectively, lead to the following
equivalent system
$$
\bigg[\partial_r+\frac{1}{2r}-\frac{i}{r}\partial_\theta\bigg]x_2-ihv y_1=i\lambda x_1,\qquad
\bigg[\partial_r+\frac{1}{2r}+\frac{i}{r}\partial_\theta\bigg] x_1-i hv y_2=i\lambda x_2,
$$
$$
\bigg[ \partial_r+\frac{1}{2r}-\frac{i}{r}\partial_\theta\bigg]y_2+ihv x_1=-i\lambda y_1,\qquad
\bigg[ \partial_r+\frac{1}{2r}+\frac{i}{r}\partial_\theta\bigg] y_1+ihv x_2=-i\lambda y_2.
$$
Here $x_1=e^{i\theta} \chi_1$,  $x_2=\chi_2$, $y_1=\eta_1$, $y_2=e^{-i\theta}\eta_2$.
From here it is found, by introducing the variables $U=x_2+y_2$, $V=x_1-y_1$, $W=x_2-y_2$ and $X=x_1+y_1$ and adding and subtracting the equations properly, the decoupled in pair system
$$
\bigg[\partial_r+\frac{1}{2r}-\frac{i}{r}\partial_\theta\bigg]U=i(\lambda-hv) V,\qquad 
\bigg[\partial_r+\frac{1}{2r}+\frac{i}{r}\partial_\theta\bigg]V=i(\lambda+hv) U,
$$
\be\lb{desacople}
\bigg[\partial_r+\frac{1}{2r}-\frac{i}{r}\partial_\theta\bigg]W=i(\lambda+hv) X,\qquad
\bigg[\partial_r+\frac{1}{2r}+\frac{i}{r}\partial_\theta\bigg] X=i(\lambda-hv) W.
\ee
The reader can check that compatibility condition for both decoupled systems reduces, after the Fourier expansion $U=\sum_{n=-\infty}^\infty U_{n} \;e^{i n\theta}$ and the analogous for $V$, $W$ and $X$,
to Bessel equations of half-integral order for the Fourier components. Furthermore, the standard recursion formulas for these Bessel solutions allows to find the full solution from (\ref{desacople}). In the core where $B(r)=F(r)=0$ the equations are decoupled directly,
they are simpler than for the outside zone and also are related to Bessel functions. The resulting expressions can be worked out in straightforward manner and, after going back to the original variables $\chi_i$ and $\eta_i$ with $i=1,2$, the result are the formulas (\ref{abovo}) for the region outside and 
\be\lb{generation1}
\chi_1=\sum_{n=-\infty}^\infty e^{in\theta} a_{n+1} J_{n}(\lambda\;r),\qquad
\chi_2=\sum_{n=-\infty}^\infty e^{in\theta} i a_{n} J_{n}(\lambda\;r).
\ee
\be\lb{generation2}
\eta_1= \sum_{n=-\infty}^\infty e^{in\theta}b_{n+1} J_{n}(\lambda\;r),\qquad
\eta_2=- \sum_{n=-\infty}^\infty e^{in\theta}  i b_{n} J_{n}(\lambda\;r),
\ee
for the inside region. The unknown coefficients $a_n$ and $b_n$ and the ones $c_{in}$ and $d_{in}$ with $i=1,2$ in (\ref{abovo}) have to be determined by the continuity of the solution at $r=r_0$, leading to the system (\ref{conti}) described above.
Note that the Bessel functions $Y_n(x)$ are discarded in the inner region, due to the fact that they are highly divergent at $x=0$. The generalization of these formulas for arbitrary winding number can be done straightforwardly.


\begin{thebibliography}{99}
\bibitem{witten} E. Witten Nucl. Phys. B 249 (1985) 557.
\bibitem{escenario1} J. P. Ostriker, A. C. Thompson, and E. Witten Phys. Lett. B 180 (1986) 231.
\bibitem{escenario2} R. Jeannerot, Phys. Rev. Lett. 77 (1996)
3292.
\bibitem{escenario3} G. Lazarides and Q. Shafi Phys. Lett. B
151 (1985) 123.
\bibitem{escenario4} N. Ganoulis and G. Lazarides, Nucl.
Phys. B 316 (1989) 443.
\bibitem{escenario5} A. Iwazaki, Phys. Rev. Lett. 79 (1997) 2927.
\bibitem{escenario7}Y. Abe, Y. Hamada, and K. Yoshioka, JHEP 06 (2021) 172.
\bibitem{escenario8} P. Agrawal, A. Hook, J. Huang, and G. Marques-Tavares JHEP 01 (2022) 103.
\bibitem{vorton1} R. L. Davis and E. P. S. Shellard Nucl. Phys. B 323 (1989)
209.
\bibitem{vorton2} R. L. Davis and E. P. S. Shellard
Phys. Lett. B 209 (1988) 485.
\bibitem{vorton3} R. H. Brandenberger, B. Carter, A.C. Davis, and M. Trodden Phys. Rev. D 54 (1996) 6059.
\bibitem{vorton33} B. Carter and A.-C. Davis
Phys.Rev. D61 (2000) 123501.
\bibitem{vorton4} C. J. A. P. Martins and E. P. S. Shellard Phys.
Lett. B 445 (1998) 43.
\bibitem{peter}  X. Martin and P. Peter Dynamical Phys.Rev. D51 (1995) 4092.
\bibitem{vorton5} C. J. A. P. Martins and E. P. S. Shellard  Phys. Rev. D 57
(1998) 7155.
\bibitem{ringeval} P. Peter and C. Ringeval
JCAP 05 (2013) 005.

\bibitem{ringeval2} Pierre Auclair, Patrick Peter, Christophe Ringeval and Daniele Steer JCAP 2103 (2021) 098.

\bibitem{vorton6} A. Cordero-Cid, X. Martin, and P. Peter Phys. Rev. D 65
(2002) 083522.
\bibitem{gangui} Alejandro Gangui, Patrick Peter, Celine Boehm Phys.Rev. D57 (1998) 2580.

\bibitem{vorton9}  M. Ibe, S. Kobayashi, Y. Nakayama, and S. Shirai JHEP 05 (2021) 217.
\bibitem{vorton99} H. Fukuda, A. Manohar, H. Murayama and O. Telem J. High Energ. Phys. 2021 (2021) 52.
\bibitem{davis} Anne-Christine Davis, Warren B. Perkins  Phys. Lett. B 393 (1997) 46.

\bibitem{rayos1} Silvano Bonazzola and Patrick Peter Astropart.Phys.7 (1997) 161.
\bibitem{rayos2} Luis Masperi and Guillermo Silva  Astropart. Phys. 8 (1998) 173.
\bibitem{rayos3} Luis Masperi and Milva Orsaria Int. J. Mod. Phys. A 14 (1999) 3581.
\bibitem{kopnin0} E. B. Sonin, Rev. Mod. Phys. 59 (1987) 87.
\bibitem{kopnin00} N. B. Kopnin, Theory of Nonequilibrium Superconductivity
(Oxford University Press, 2001).
\bibitem{kopnin1} N. B. Kopnin, Rep. Prog. Phys. 65 (2002) 1633.
\bibitem{kopnin2} E. B. Sonin, in Vortices in Unconventional Superconductors and Superfluids, edited by R. P. Huebener,
N. Schopohl, and G. E. Volovik (Springer-Verlag, 2002)
pp. 119–145.
\bibitem{kopnin3} A. L. Shelankov, in Vortices in Unconventional Superconductors and Superfluids, edited by R. P. Huebener,
N. Schopohl, and G. E. Volovik (Springer-Verlag, 2002)
pp. 147—166.
\bibitem{kopnin3} G. E. Volovik, The Universe in a Helium Droplet (Oxford
University Press, 2003).
\bibitem{kopnin4} M. Stone, Phys. Rev. B 54 (1996)13222.

\bibitem{kopnin5} H. Suhl, Phys. Rev. Lett. 14 (1965) 226.
\bibitem{kopnin6} V. N. Popov, Zh. Eksp. Teor. Fiz 64 (1973) 672 [Sov.
Phys.–JETP 37, 341 (1973)].
\bibitem{kopnin7} N. B. Kopnin, Pis’ma Zh. Eksp. Teor. Fiz. 27 (1978) 417,
[JETP Lett., 23, 578 (1976)].
\bibitem{kopnin8} G. Baym and E. Chandler, J. Low Temp. Phys. 50 (1987) 57.
\bibitem{kopnin9} J.-M. Duan and A. J. Leggett, Phys. Rev. Lett. 68, 1216
(1992), er., ibid 69, 1148.
\bibitem{kopnin10} J.-M. Duan, Phys. Rev. B 48 (1993) 333.
\bibitem{kopnin11} J.-M. Duan, Phys. Rev. B 49 (1994) 12381.
\bibitem{kopnin12} E. B. Sonin, V. B. Geshkenbein, A. van Otterlo, and
G. Blatter, Phys. Rev B 57 (1998) 575.
\bibitem{kopnin13} N. B. Kopnin and V. M. Vinokur, Phys. Rev. Lett. 81,
3952 (1998).
\bibitem{kopnin14} G. E. Volovik, Pis’ma Zh. Eksp. Teor. Fiz. 57 (1993) 233,
[JETP Letters, 57, 244 (1993)].
\bibitem{kopnin15} N. B. Kopnin and V. E. Kravtsov, Pis’ma Zh. Eksp. Teor.
Fiz. 23, (1976) 631 [JETP Lett., 23, 578 (1976)].
\bibitem{kopnin16} E. B. Sonin, Zh. Eksp. Teor. Fiz. 69 (1975) 921 [Sov.
Phys.–JETP, 42, 469 (1976)].
\bibitem{kopnin17} N. B. Kopnin and V. E. Kravtsov, Zh. Eksp. Teor. Fiz. 71,
(1976) 1664 [Sov. Phys.–JETP, 44, 861 (1976)].
\bibitem{kopnin18} Y. M. Gal’perin and E. B. Sonin, Fiz. Tverd. Tela
(Leningrad) 18 (1976) 3034, [Sov. Phys.–Solid State 18,
1768 (1976)].
\bibitem{kopnin19} E. B. Sonin, Phys. Rev. B 55, (1997) 485.
 \bibitem{kopnin20} G. E. Volovik, Pis’ma Zh. Eksp. Teor. Fiz. 67 (1998) 502,
[JETP Lett., 67, 528 (1998)].
\bibitem{pelea1} E. B. Sonin  Phys. Rev. B 87 (2013) 134515.
\bibitem{kopnin20} F. Gaitan  Phys. Rev. B 51 (1985) 9061.
\bibitem{kopnin21} W. E. Goff, F. Gaitan and M. Stone, Phys. Lett. A 136 (1989) 433.
\bibitem{kopnin22} A. Garg, V. P. Nair and M. Stone, Ann. Phys. 173 (1987) 149.
\bibitem{kopnin23} F. Gaitan, Phys. Lett. A 151 (1991) 551.
\bibitem{kopnin24} P. Ao and D. J. Thouless, Phys. Rev. Lett. 70 (1993) 2158.
\bibitem{kopnin25} P. Ao, Q. Niu and D. J. Thouless, Physica B 194-196 (1994) 1453.
\bibitem{pelea2} D.J. Thouless, P. Ao and Q. Niu, Phys. Rev. Lett. 76 (1996) 3758.
\bibitem{kopnin26} G.E. Volovik, JETP Lett. 44 (1986) 185.
\bibitem{kopnin27}  F. Gaitan, Phys. Lett., A 178 (1993) 449.
\bibitem{kopnin28}  G.E. Volovik, JETP 75 (1992) 990.
\bibitem{kopnin29}  N.B. Kopnin, Physica B 210 (1995) 267.
\bibitem{pelea3} G. Volovik Phys. Rev. Lett. 77 (1996) 4687.
\bibitem{experimento} T. D. C. Bevan, A. J. Manninen, J. B. Cook, A. J. Armstrong, J. R. Hook, and H. E. Hall Phys. Rev. Lett. 74, 750–753 (1995); 74, 3092 (E) (1995).
\bibitem{goldstone} Jeffrey Goldstone and Frank Wilczek
Phys. Rev. Lett. 47 (1981) 986.

\bibitem{GrRy}
I.~S. Gradshteyn and I.~M. Ryzhik.
\newblock {\em Table of integrals, series, and products}.
\newblock Elsevier/Academic Press, Amsterdam, seventh edition, 2007.
\newblock Translated from the Russian, Translation edited and with a preface by
  A. Jeffrey and D. Zwillinger.

\bibitem{taylor}
M.~Taylor.
\newblock Bessel functions and {H}ankel transforms.
\newblock
  {\footnotesize\url{https://mtaylor.web.unc.edu/wp-content/uploads/sites/16915/2018/04/bessel2.pdf}},
  2018.
  \bibitem{ruijsenaars} N. Ruijsenaars, Annals of Physics  146 (1983) 1.
  \bibitem{alford} M. G. Alford and F. Wilczek, Phys. Rev. Lett. 62 (1989) 1071.
  \bibitem{vera} F. Vera and I. Schmidt, Phys. Rev. D 42 (1990) 3591.
  \bibitem{zeromode}  S. C. Davis, W. B. Perkins and A. C. Davis, Phys. Rev. D 62 (2000)
043503.
\bibitem{zeromode2} R. Jackiw and P. Rossi Nucl. Phys. B 190 (1981) 681.
\bibitem{CoHi}
R.~Courant and D.~Hilbert.
\newblock {\em Methods of mathematical physics. {V}ol. {I}}.
\newblock Interscience Publishers, Inc., New York, N.Y., 1953.

\bibitem{duren}
P.~Duren.
\newblock {\em Harmonic mappings in the plane}, volume 156 of {\em Cambridge
  Tracts in Mathematics}.
\newblock Cambridge University Press, Cambridge, 2004.


\bibitem{JaRo}
R.~Jackiw and P.~Rossi.
\newblock Zero modes of the vortex-fermion system.
\newblock {\em Nuclear Physics B}, 190(4):681--691, 1981.

\bibitem{nohl}
C.~R. Nohl.
\newblock Bound-state solutions of the {D}irac equation in extended hadron
  models.
\newblock {\em Phys. Rev. D}, 12:1840--1842, Sep 1975.

\bibitem{rubakov}
V.~Rubakov.
\newblock {\em Classical theory of gauge fields}.
\newblock Princeton University Press, Princeton, NJ, 2002.
\newblock Translated from the 1999 Russian original by S. S. Wilson.


\end{thebibliography}
\end{document}